\documentclass[prb,twocolumn,preprintnumbers,amsmath,amssymb]{revtex4}% Physical Review B

\usepackage{graphicx}% Include figure files
\graphicspath{{fig/}}
\usepackage{dcolumn}% Align table columns on decimal point\varphi 
\usepackage{bm}% bold math
\usepackage{amsmath,amssymb}
\usepackage[T1]{fontenc}
\usepackage{textcomp}
\usepackage{color}
\begin{document}

\title{Magnonic topological insulators in antiferromagnets 
%and Wiedemann-Franz law
}

\author{Kouki Nakata,$^1$ Se Kwon Kim,$^2$ Jelena Klinovaja,$^1$ and Daniel Loss$^1$}

\affiliation{$^1$Department of Physics, University of Basel, Klingelbergstrasse 82, CH-4056 Basel, Switzerland   \\
$^2$Department of Physics and Astronomy, University of California, Los Angeles, California 90095, USA  
}

\date{\today}

\begin{abstract}
Extending the notion of symmetry protected topological phases to insulating antiferromagnets (AFs) described in terms of opposite magnetic dipole moments associated with the magnetic N$\acute{{\rm{e}}} $el  order, we establish a bosonic counterpart of  topological insulators in semiconductors.
Making use of the Aharonov-Casher effect, induced by electric field gradients, we  propose a magnonic analog of the quantum spin Hall effect (magnonic QSHE) for edge states that carry  helical magnons.
We show that such up and down magnons form the same Landau levels and perform cyclotron motion with the same frequency but propagate in opposite direction.
The insulating AF becomes characterized by a  topological ${\mathbb{Z}}_{2}$  number consisting of the Chern integer associated with each helical magnon edge state.
Focusing on the  topological Hall phase for magnons, we study bulk magnon effects such as magnonic spin, thermal, Nernst, and Ettinghausen effects, as well as the thermomagnetic properties of helical magnon transport both in topologically trivial and nontrivial bulk AFs and establish the magnonic Wiedemann-Franz law.
We show that our predictions are within experimental reach with current device and measurement techniques.
\end{abstract}

%\pacs{75.30.Ds, 73.43.-f, 75.47.-m, 77.55.Nv, 85.75.-d,72.25.-b, 75.85.+t}

\maketitle

%%%%%%%%%%%%%%%%%
\section{Introduction}
\label{sec:Intro}
%%%%%%%%%%%%%%%%%

\begin{figure}[t]
\begin{center}
\includegraphics[width=7.5cm,clip]{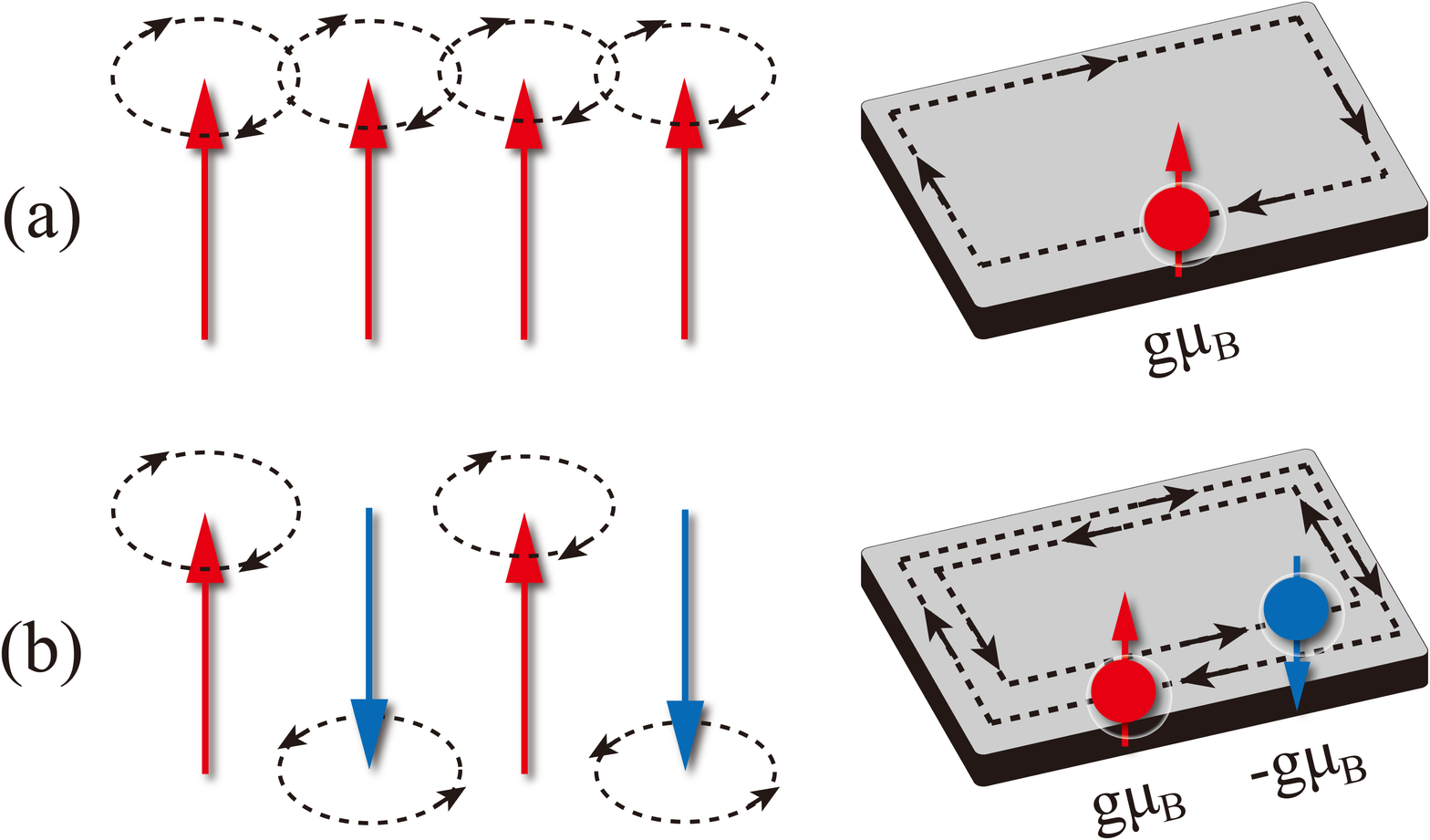}
\caption{(Color online)
Left: Schematic representation of spin excitations in 
(a) a FM with the uniform ground state magnetization and (b) an AF with classical magnetic N$\acute{{\rm{e}}} $el  order.
Right: Schematic representation of edge magnon states in a two-dimensional topological (a) FM and (b) AF.
(a) Chiral edge magnon state where magnons with a magnetic dipole moment $ g\mu_{\rm{B}}  {\bf e}_z$ propagate along the edge of a finite sample in a given direction.
(b) Helical edge magnon state where up and down magnons ($\sigma =\pm 1$) with opposite magnetic dipole moments $\sigma  g\mu_{\rm{B}}  {\bf e}_z$ propagate along the edge in opposite directions. The AF thus forms a bosonic analog of a TI characterized by two edge modes with opposite chiralities and can be identified with two independent copies (with opposite magnetic moments) of  single-layer FMs shown in (a).
}
\label{fig:HelicalAFChiralFM}
\end{center}
\end{figure}

Since the observation of quasiequilibrium Bose-Einstein condensation \cite{demokritov} of magnons in an insulating ferromagnet (FM) at room temperature, 
the last decade has seen remarkable and rapid development of a new branch of magnetism, dubbed magnonics \cite{MagnonSpintronics,magnonics,ReviewMagnon}, aimed at utilizing magnons, the quantized version of spin-waves, as  substitute for electrons with the advantage of low-dissipation. Magnons are chargeless bosonic quasi-particles with a magnetic dipole moment $g \mu_{\rm{B}} {\mathbf{e}}_z$  that can serve as a carrier of information in units of the Bohr magneton $\mu_{\rm{B}}$.
In particular, insulating FMs \cite{spinwave,onose,WeesNatPhys,MagnonHallEffectWees} that possess a macroscopic magnetization [Fig. \ref{fig:HelicalAFChiralFM} (a)] have been playing an essential role in magnonics.
Spin-wave spin current \cite{spinwave,WeesNatPhys}, thermal Hall effect of magnons \cite{onose}, and Snell's law \cite{Snell_Exp,Snell2magnon} for spin-waves have been experimentally established and just this year the magnon planar Hall effect \cite{MagnonHallEffectWees} has been observed.
A  magnetic dipole moving in an electric field acquires a geometric phase by the Aharonov-Casher \cite{casher,Mignani,magnon2,KKPD,ACatom,AC_Vignale,AC_Vignale2} (AC) effect, which is analogous to the Aharonov-Bohm effect \cite{bohm,LossPersistent,LossPersistent2}  of electrically charged particles in magnetic fields, and the AC effect in magnetic systems has  also been experimentally confirmed \cite{ACspinwave}.

Under a strong magnetic field, two-dimensional electronic systems can exhibit the integer quantum Hall effect~\cite{QHEcharge} (QHE), which is characterized by chiral edge modes. Thouless, Kohmoto, den Nijs, and Nightingale~\cite{TKNN,Kohmoto} (TKNN) described the QHE ~\cite{AFQHE,HaldaneQHEnoB,VolovikQHE,JK_IFQAHE} in terms of a topological invariant, known as TKNN integer, associated with bulk wave functions \cite{HalperinEdge,BulkEdgeHatsugai} in  momentum space. This introduced the notion of topological phases of matter, which has been attracting much attention over the last decade. In particular, in 2005, Kane and Mele~\cite{Z2topo, QSHE2005} have shown that graphene in the absence of a magnetic field exhibits a quantum spin Hall effect (QSHE)~\cite{QSHE2005,QSHE2006,TIreview,TIreview2}, which is characterized by a pair of gapless spin-polarized edge states. These helical edge states are protected from backscattering by  time-reversal symmetry (TRS), forming in this sense topologically protected Kramers pairs.
This can be seen to be the first example of a symmetry protected topological (SPT) phase \cite{SPTreviewXGWen,SPTreviewSenthil,SenthilLevin} and it is now classified as a  topological insulator (TI)~\cite{Z2topo,Z2topoHaldane,Z2SM,TIreview,TIreview2}, which is characterized by a ${\mathbb{Z}}_{2}$ number, as  the TKNN integer \cite{TKNN,Kohmoto} associated with each edge state.

In this paper we extend the notion of topological phases to insulating antiferromagnets (AFs) in the N$\acute{{\rm{e}}} $el ordered phases  which do not possess a macroscopic magnetization, see Fig. \ref{fig:HelicalAFChiralFM} (b). The component  of the total spin along the N$\acute{{\rm{e}}} $el vector is assumed to be conserved, and it is this conservation law which plays the role of the TRS (which is broken in the ordered AF) that protects the topological phase and helical edge states against nonmagnetic impurities and the details \footnote{We assume a sample in the absence of magnetic disorder that breaks the global spin rotation symmetry about the $z$ axis such as the uncompensated surface magnetization.} of the surface. \cite{SurfaceMode} 
In particular, using magnons \cite{AndersonAF,RKuboAF,AFspintronicsReview2,AFspintronicsReview} we thus establish a bosonic counterpart of the TI and propose a magnonic QSHE  resulting from the  N$\acute{{\rm{e}}} $el order in AFs.

In Ref. [\onlinecite{KJD}], motivated by the above-mentioned remarkable progress in recent experiments, we \cite{magnon2,magnonWF,ReviewMagnon,KevinHallEffect} have  proposed a way to electromagnetically realize the `quantum' Hall effect of magnons in FMs, in the sense that the magnon Hall conductances are characterized by a Chern number \cite{TKNN,Kohmoto} in an almost flat magnon band, which hosts a chiral edge magnon state, see Fig. \ref{fig:HelicalAFChiralFM} (a). \footnote{See also Refs. [\onlinecite{SMreviewMagnon,Haldane2,katsura,SKK_HKM}] for topological aspects of magnons in FMs, Ref. [\onlinecite{TopoMagBandKagome}] for observation of a topological magnon band \cite{KJD,KevinHallEffect,RSdisorder}, and Refs. [\onlinecite{PhotonTopo,PhotonTopo3D,HaldanePhoton}] for photonic TIs.}
By providing a topological description \cite{NiuBerry,Kohmoto,TKNN} of the classical magnon Hall effect induced by the AC effect, which was proposed in Ref. [\onlinecite{magnon2}], we developed it further into the magnonic `quantum' Hall effect and appropriately defining the thermal conductance for bosons, we found that the magnon Hall conductances in such topological FMs obey a Wiedemann-Franz \cite{WFgermany} (WF) law for magnon transport \cite{magnonWF,ReviewMagnon}. 
%%%%%%%%%%%%%%%%%%%%%
In this paper, motivated by the recent experimental \cite{SekiAF} demonstration of thermal generation of spin currents in AFs using the spin Seebeck effect \cite{uchidainsulator,ishe,ohnuma,adachi,adachiphonon,xiao2010,OnsagerExperiment,Peltier} and by the report \cite{MagnonNernstAF,MagnonNernstAF2,MagnonNernstExp} of the magnonic spin Nernst effect in AFs, 
we develop Ref. [\onlinecite{KJD}] further into the AF regime~\cite{AFspintronicsReview2,AFspintronicsReview,RKuboAF,AndersonAF,Kevin2,DLquantumAF} and propose a magnonic analog of the QSHE \cite{QSHE2005,QSHE2006,Z2topo,Z2topoHaldane,Z2SM,TIreview,TIreview2} for edge states that carry helical edge magnons [Fig. \ref{fig:HelicalAFChiralFM} (b)] due to the AC phase.
Focusing on helical magnon transport both in topologically trivial and nontrivial bulk \cite{HalperinEdge,BulkEdgeHatsugai} AFs,
we also study thermomagnetic properties and discuss the universality of the magnonic WF law \cite{magnonWF,KJD,ReviewMagnon}.
Using magnons in insulating AFs characterized by the N$\acute{{\rm{e}}} $el magnetic order, we thus establish the bosonic counterpart of TIs \cite{Z2topo,Z2topoHaldane,Z2SM,TIreview,TIreview2}.

At sufficiently low temperatures, effects of magnon-magnon and magnon-phonon interactions become \cite{magnonWF,adachiphonon,Tmagnonphonon} negligibly small.
Indeed, Ref. [\onlinecite{Tmagnonphonon}] reported measurements of magnets at low temperature $T \lesssim {\cal{O}}(1) $K, where the exponent of the temperature dependence of the phonon thermal conductance is larger than the one for magnons. 
This indicates that for thermal transport the contribution of phonons dies out more quickly than that of magnons with decreasing temperature.
We then focus on noninteracting magnons at such low temperatures throughout this work  \footnote{Within the mean-field treatment, interactions between magnons works \cite{magnonWF} as an effective magnetic field and the results qualitatively remain identical. A certain class of QHE in systems with interacting bosons, a SPT phase \cite{SPTreviewXGWen,SPTreviewSenthil} is implied in Refs.~[\onlinecite{SenthilLevin}] and [\onlinecite{SPTcoldatom}].}
and assume that the total spin along the $z$ direction is conserved and thus remains a good quantum number.

This paper is organized as follows.
In Sec. \ref{sec:trivial} we introduce the model system for magnons with a quadratic dispersion because of an easy-axis spin anisotropy in a topologically trivial bulk AF and find that the dynamics can be described as the combination of two independent copies of that in a FM and thus derive the identical magnonic WF law for a topologically trivial FM and AF.
%%%%%%%
In Sec. \ref{subsec:ACmagnon}, introducing the model system for magnons in the presence of an AC phase induced by an electric field gradient, we find the correspondence between the single magnon Hamiltonian and the one of an electrically charged particle moving in a magnetic vector potential, and see that the force acting on magnons is invariant under a gauge transformation.
%%%%%%%
In Sec. \ref{subsec:MQSHE}, we see that each magnon with opposite magnetic dipole moment inherent to the N$\acute{{\rm{e}}} $el  order forms the same Landau levels and performs cyclotron motion with the same frequency but in opposite direction, leading to the helical edge magnon state. We find that the AF in the topological ${\mathbb{Z}}_{2}$ phase is characterized by a Chern number associated with each edge state.
%%%%%%%
In Sec. \ref{subsec:LatticeNumCal}, introducing a tight-binding representation (TBR) of the magnon Hamiltonian, we obtain the magnon energy spectrum and helical edge states numerically, for constant and periodic electric field gradients.
%%%%%%%
In Sec. \ref{subsec:TopoHallBulk}, we study thermomagnetic properties of  Hall transport of bulk magnons, with focus on the  the helical edge magnon states, and analyze the differences between the topological and non-topological phases of the AF.
In Sec. \ref{sec:experimentAF} we give some concrete estimates for experimental candidate materials. 
Finally, we summarize and give some conclusions in Sec. \ref{sec:sum}, and remark open issues in Sec. \ref{sec:discussion}. Technical details are deferred to the Appendices.

%%%%%%%%%%%%%%%%%%
\section{Topologically trivial AF}
\label{sec:trivial}
%%%%%%%%%%%%%%%%%%

In this section, we consider a topologically trivial AF on a three-dimensional ($d=3$) cubic lattice in the ordered phase
with the N$\acute{{\rm{e}}} $el  order parameter along the $z$ direction, see Fig. \ref{fig:HelicalAFChiralFM} (b).
Spins of length $S$ on each bipartite sublattice, denoted by A and B, satisfy \cite{altland,MagnonNernstAF,MagnonNernstAF2} 
$  {\mathbf{S}}_{\rm{A}}  =  - {\mathbf{S}}_{\rm{B}}= S {\mathbf{e}}_z    $ in the ground state.
The magnet is described by the following spin Hamiltonian, \cite{RKuboAF,AndersonAF}
\begin{eqnarray}
 {\cal{H}} = J \sum_{\langle ij \rangle}  {\mathbf{S}}_i  \cdot  {\mathbf{S}}_j -\frac{\cal{K}}{2} \sum_{i} (S_i^z)^2,
\label{eqn:H} 
\end{eqnarray}
where $J >0$ parametrizes the antiferromagnetic exchange interaction between the nearest-neighbor spins and ${\cal{K}} >0$ is the easy-axis anisotropy \cite{CrOexp,CrOexp2} that ensures the magnetic N$\acute{{\rm{e}}} $el  order along the $z$ direction. Since the Hamiltonian is invariant under global spin rotations about the $z$ axis, the $z$ component of the total spin is a good quantum number (i.e. conserved). Therefore, magnons, quanta of spin waves, have well-defined spin along the $z$ axis, as will be shown explicitly below. Using the sublattice-dependent Holstein-Primakoff~\cite{HP,altland,MagnonNernstAF,MagnonNernstAF2} transformation,
$ S_{i{\rm{A}}}^+ = \sqrt{2S}[1-a_i^\dagger a_i / (2S)]^{1/2} a_i$, 
$S_{i{\rm{A}}}^z = S - a_i^\dagger a_i$, 
$ S_{j{\rm{B}}}^+ = \sqrt{2S}[1-b_j^\dagger b_j / (2S)]^{1/2} b_j^{\dagger }$, 
$S_{j{\rm{B}}}^z = - S + b_j^\dagger b_j$, 
the spin degrees of freedom in Eq. (\ref{eqn:H}) can be recast in terms of bosonic operators that satisfy the commutation relations,
$[a_{i}, a_{j}^{\dagger }] = \delta_{i,j}$ and $[b_{i}, b_{j}^{\dagger }] = \delta_{i,j}$ and all other commutators  between the annihilation (creation) operators $a_{i}^{(\dagger)}$ and $b_{i}^{(\dagger)}$ vanish to the lowest order in $1/S$, assuming large spins $S \gg  1$.
Further performing the well-known Bogoliubov transformation \cite{altland} (see Appendix \ref{sec:Mchirality} for details), the system can be mapped onto a system of non-interacting spin-up and spin-down magnons, which carry opposite magnetic dipole moment~\cite{altland} $\sigma  g\mu_{\rm{B}}  {\bf e}_z$ with $\sigma = \pm 1$.
The transformed Hamiltonian assumes diagonal form \cite{RKuboAF,AndersonAF},
$ {\cal{H}} =  \sum_{\mathbf{k}}  \hbar \omega _{\mathbf{k}} ({\cal{A}}_{\mathbf{k}}^{\dagger } {\cal{A}}_{\mathbf{k}} + {\cal{B}}_{\mathbf{k}}^{\dagger } {\cal{B}}_{\mathbf{k}})$, in terms of Bogoliubov quasi-particle operators, ${\cal{A}}_{\mathbf{k}}$ and ${\cal{B}}_{\mathbf{k}}$, satisfying $ [{\cal{A}}_{\mathbf{k}}, {\cal{A}}_{{\mathbf{k^{\prime}}}}^{\dagger }]= \delta_{{\mathbf{k}}, {\mathbf{k^{\prime}}}}  $
and $ [{\cal{B}}_{\mathbf{k}}, {\cal{B}}_{{\mathbf{k^{\prime}}}}^{\dagger }]= \delta_{{\mathbf{k}}, {\mathbf{k^{\prime}}}}  $ with all the other commutators vanishing.
%%%%%%%%%%%%%%%%%%%
Within the long wave-length approximation and assuming a spin anisotropy \cite{CrOexp,CrOexp2} at low temperature,
the dispersion \cite{RKuboAF} becomes gapped and parabolic \footnote{As long as the temperature is lower than the magnon gap $\Delta$, thermomagnetic and topological properties  remain qualitatively the same also for magnons with a linear dispersion \cite{KJD,magnonWF}.}
in terms of $k = | {\mathbf{k}} |  $, $\hbar \omega _{{\mathbf{k}}} = Dk^2 + \Delta$, where $   D = JSa^2/\sqrt{\kappa^2 + 2\kappa}  $ 
 parametrizes the inverse of the `magnon mass' (see below), 
$ \Delta  = 2dJS\sqrt{\kappa^2 + 2\kappa} $ is the magnon gap, $ \kappa = {\cal{K}}/(2dJ)   $, $a$ denotes the lattice constant, and $d=3$ is the dimension of the cubic lattice ({\it{e.g.}}, $d=2$ for the square lattice).
Note that the dispersion becomes linear $\hbar \omega _{{\mathbf{k}}} \propto  k $ in the absence of the spin anisotropy \cite{AndersonAF}, see Appendix \ref{sec:Mchirality} for details.

Since the $z$ component of the total spin is given by \cite{MagnonNernstAF}
$  S^z = \sum_{i} (S_{i{\rm{A}}}^z  + S_{i{\rm{B}}}^z) 
= \sum_{\mathbf{k}} (-a_{\mathbf{k}}^{\dagger } a_{\mathbf{k}} + b_{\mathbf{k}}^{\dagger } b_{\mathbf{k}})
= \sum_{\mathbf{k}}(-{\cal{A}}_{\mathbf{k}}^{\dagger } {\cal{A}}_{\mathbf{k}} + {\cal{B}}_{\mathbf{k}}^{\dagger } {\cal{B}}_{\mathbf{k}}) $,
the ${\cal{A}}$ (${\cal{B}}$) magnon carries $\sigma = -1$ ($+1$) spin angular momentum along the $z$ direction
and can be identified with a down (up) magnon.
Thus the low-energy magnetic excitation of the AF [Eq.~(\ref{eqn:H})] can be described as chargeless bosonic quasiparticles carrying a magnetic dipole moment $\sigma  g\mu_{\rm{B}}  {\bf e}_z$ with $\sigma = \pm 1$ [Fig. \ref{fig:HelicalAFChiralFM} (b)], where $g$ is the $g$-factor of the constituent spins and $\mu_{\rm{B}}$ is the Bohr magneton. 
Throughout this paper, we work under the assumption that the total spin along the $z$ direction is conserved and remains a good quantum number.

In the presence of an external magnetic field $B\geq 0$ along the $z$ axis $ {\mathbf{B}} = B {\mathbf{e}}_z  $, 
the degeneracy is lifted and the low-energy physics of the AF at sufficiently low temperatures where effects of magnon-magnon and magnon-phonon interactions become \cite{magnonWF,adachiphonon,Tmagnonphonon} negligibly small is described by the Hamiltonian
\begin{eqnarray}
{\cal{H}} =  \sum_{\sigma = \uparrow , \downarrow } \sum_{\mathbf{k}}  \hbar \omega _{{\mathbf{k}}\sigma } a_{{\mathbf{k}}\sigma }^{\dagger } a_{{\mathbf{k}}\sigma }, 
\label{eqn:H2} 
\end{eqnarray}
where $\sigma = \uparrow $ and $\sigma = \downarrow$ denote the up magnon ($\sigma =1$) and the down magnon ($\sigma = -1$), respectively. Here, $\hbar \omega _{{\mathbf{k}}\sigma } = Dk^2 + \Delta_{\sigma } $ and $\Delta_{\sigma } = \Delta - \sigma g \mu _{\rm{B}} B $ are the energy and the gap of spin-$\sigma$ magnons; $a_{{\mathbf{k}} \sigma }^{\dagger } a_{{\mathbf{k}} \sigma }$ is the number operator of spin-$\sigma$ magnons.
Throughout the paper, we adopt the aforementioned notations for simplicity.
We consider a magnetic field that is much weaker than the anisotropy, i.e., $ g \mu _{\rm{B}} B \ll  \Delta $,
where the spin anisotropy prevents spin flop transition.

%%%%%%%%%%%%%%%%%%%%
\subsection{Onsager coefficients}
\label{subsec:Onsager}
%%%%%%%%%%%%%%%%%%%%

The two magnon modes, up and down, are completely decoupled in the AF described by Eq. (\ref{eqn:H2}).
Therefore the dynamics of magnons in the AF can be described as the combination of two independent copies of the dynamics of magnons in a FM for each mode $\sigma =\pm 1$.
For spin-$\sigma$ magnons, a magnetic field gradient $ \partial _x B $ along the $x$ axis works as a driving force ${\mathbf{F}}_{B} = F_{\sigma} {\mathbf{e}}_x$ with $F_{\sigma} = \sigma g \mu _{\rm{B}} \partial _x B $. Since the directions of the force are opposite for the two magnon modes $\sigma = \pm 1$, the magnetic field gradient generates {\it{helical}} magnon transport in the topologically trivial bulk AF, Eq. (\ref{eqn:H2}), as will be shown explicitly below.
Specifically, the magnetic field and temperature gradients generate magnonic spin and heat currents, $j_{x \sigma }$ and $ j_{x\sigma }^Q $, respectively, along the $x$ direction. 
Within the linear response regime, each Onsager coefficient $L_{ij \sigma }$ ($i, j = 1, 2$ ) is defined by
\begin{eqnarray}
%%%%%%%%%%%%%%%%%%%%%%%%%%%%%
\begin{pmatrix}
\langle j_{x \sigma }  \rangle  \\  \langle j_{x\sigma }^Q \rangle
\end{pmatrix}
%%%%%%%%%%%%%%%%%%%%%%%%%%%%%
=
%%%%%%%%%%%%%%%%%%%%%%%%%%%%%
\begin{pmatrix}
L_{11 \sigma } & L_{12\sigma } \\ L_{21\sigma } & L_{22\sigma }
\end{pmatrix}
%%%%%%%%%%%%%%%%%%%%%%%%%%%%%
%%%%%%%%%%%%%%%%%%%%%%%%%%%%%
\begin{pmatrix}
  \partial _x B  \\   - \partial _x T/T
\end{pmatrix}.
%%%%%%%%%%%%%%%%%%%%%%%%%%%%%
\label{eqn:2by2af}
\end{eqnarray}
A straightforward calculation using the Boltzmann equation \cite{mahan,Basso,Basso2,Basso3} gives the following coefficients (see Appendix \ref{sec:Boltzmann} for details),
\begin{subequations}
\begin{align}
{{ L}}_{11\sigma } &= (g \mu _{\rm{B}})^2 {\cal{C}}  \cdot  {\rm{Li}}_{3/2}({\rm{e}}^{-b_{\sigma }}),
  \label{eqn:11}  \\
%%%%%%%%%%%%%%%%%
{{ L}}_{12\sigma } &= \sigma   g \mu _{\rm{B}}  k_{\rm{B}} T {\cal{C}} 
 \Big[\frac{5}{2} {\rm{Li}}_{5/2}({\rm{e}}^{-b_{\sigma }}) +b_{\sigma }{\rm{Li}}_{3/2}({\rm{e}}^{-b_{\sigma }})  \Big]   \label{eqn:21}   \\
   &= {{ L}}_{21 \sigma },
   \label{eqn:12}    \\
%%%%%%%%%%%%%%%%
{{L}}_{22\sigma } &=    (k_{\rm{B}} T)^2 {\cal{C}} 
\Big[\frac{35}{4} {\rm{Li}}_{7/2}({\rm{e}}^{-b_{\sigma }}) +5 b_{\sigma }{\rm{Li}}_{5/2}({\rm{e}}^{-b_{\sigma }})  \nonumber   \\   
&+ b_{\sigma }^2 {\rm{Li}}_{3/2}({\rm{e}}^{-b_{\sigma }}) \Big],
 \label{eqn:22} 
%%%%%%%%%%%%%%%%%%%%%%
\end{align}
\end{subequations}
where $b_{\sigma } \equiv   \Delta_{\sigma }/(k_{\rm{B}} T)$ represents the dimensionless inverse temperature,
${\rm{Li}}_{s}(z) = \sum_{n=1}^{\infty} z^n/n^s$ is the polylogarithm function, and ${\cal{C}}  \equiv \tau  (k_{\rm{B}} T)^{3/2}/(4 \pi^{3/2} \hbar ^2 \sqrt{D})  $ with 
a phenomenologically introduced lifetime $\tau $ of magnons, which can be generated by nonmagnetic impurity scatterings and is assumed to be constant at low temperature.
The coefficients in Eq. (\ref{eqn:12}) satisfy the Onsager relation.
%%%%%%%%%%%%%%%%%%%%%%%%%%%%%%%%%
In the absence of a magnetic field, $B=0$, up and down magnons are degenerate and the degeneracy is robust against external perturbations 
due to the spin anisotropy \cite{CrOexp,CrOexp2} and the resultant magnon energy gap.
This gives $ {{ L}}_{i i \uparrow } = {{ L}}_{i i \downarrow } $, while $ {{ L}}_{i j \uparrow } = - {{ L}}_{i j \downarrow } $ for $i \not= j $ because of the opposite magnetic dipole moment $\sigma =\pm 1$.
%%%%%%%%%%%%%%%%%%%%%%%%%
Note that the particle current for each magnon $ j_{x \sigma }^{\rm{P}} $ is given by $ j_{x \sigma }^{\rm{P}} = j_{x \sigma }/(\sigma  g \mu _{\rm{B}})$,
and Eqs. (\ref{eqn:11}) and (\ref{eqn:21}) show that the magnetic field gradient generates {\it{helical}} magnon transport in the bulk AF
where magnons with  opposite magnetic moments flow in opposite $x$ directions,
while all magnons subjected to a thermal gradient flow in the same $x$ direction.
Thermomagnetic properties of such magnon transport in topologically trivial bulk AFs are summarized in Table \ref{tab:table1}.

%%%%%%%%%%%%%%%%%%%%%%%%%%%%%%%%%%%%%%%%%%%%%%%
%%%%%%%%%%%%%%%%%%%%%%%%%%%%%%%%%%%%%%%%%%%%%%%
\begin{table*}
\caption{
\label{tab:table1}
Thermomagnetic properties of magnon transport at low temperatures induced by magnetic field and thermal gradients 
both in topologically trivial and nontrivial bulk AFs, discussed in Sec. \ref{sec:trivial} and Sec. \ref{subsec:TopoHallBulk}, respectively.
The difference arises from the Chern number ${\cal{N}}_{0 \sigma } = \sigma=\pm 1 $, i.e., the relation between topological integer
$ {\cal{N}}_{0 \uparrow } + {\cal{N}}_{0 \downarrow } =0   $, while
${\cal{Z}}_{0} \equiv {({\cal{N}}_{0 \uparrow } - {\cal{N}}_{0 \downarrow })}/{2}   =   1   \   ({\rm{mod}} \, 2)$.
Here we refer to helical magnon transport where magnons with opposite magnetic dipole moments $\sigma  g\mu_{\rm{B}}  {\bf e}_z$ propagate in opposite directions. 
}
%%%%%%%%%%%%
\begin{ruledtabular}
\begin{tabular}{ccccccc}
%%%%%%%%%%%%%%%%%%%%%%%%%%%%%%%%%%%%%%%%%%%%
Antiferromagnet   & Topologically trivial bulk: Sec. \ref{sec:trivial} &  Topological bulk: Sec. \ref{subsec:TopoHallBulk}
\\ \hline   \hline
  Magnetic field gradient & Helical magnon transport  &  Magnon Hall transport \\
  %%%%%%%%%%%%%%%%%%%%%%%%%%%%%%%%%%%%%%%%%%%%
  \   & Spin: $ \sum_{\sigma } L_{11 \sigma }\not=0 $
& Spin:  $ \sum_{\sigma } L_{11 \sigma }^{yx} =0 $  \\
%%%%%%%%%%%%%%%%%%%%%%%%%%%%%%%%%%%%%%%%%%%%% 
\  & Heat: $ \sum_{\sigma } L_{21 \sigma } =0 $
& Heat:   $ \sum_{\sigma } L_{21 \sigma }^{yx} \not=0 $ \\  \hline
%%%%%%%%%%%%%%%%%%%%%%%%%%%%%%%%%%%%%%%%%%%%%
Thermal gradient & Magnon transport
& Helical magnon Hall transport  \\
%%%%%%%%%%%%%%%%%%%%%%%%%%%%%%%%%%%%%%%%%%%%%
\ & Spin: $ \sum_{\sigma } L_{12 \sigma } =0 $
&  Spin:   $ \sum_{\sigma } L_{12 \sigma }^{yx} \not=0 $ \\
%%%%%%%%%%%%%%%%%%%%%%%%%%%%%%%%%%%%%%%%%%%%%
\ & Heat: $ \sum_{\sigma } L_{22 \sigma } \not=0 $
& Heat:   $ \sum_{\sigma } L_{22 \sigma }^{yx} =0 $  \\
%%%%%%%%%%%%%%%%%%%%%%%%%%%%%%%%%%%%%%%%%%%%%
\end{tabular}
\end{ruledtabular}
\end{table*}
%%%%%%%%%%%%%%%%%%%%%%%%%%%%%%%%%%
%%%%%%%%%%%%%%%%%%%%%%%%%%%%%%%%%%

%%%%%%%%%%%%%%%%%%%%%%%%%%%
\subsection{Thermomagnetic relations}
\label{subsec:relation}
%%%%%%%%%%%%%%%%%%%%%%%%%%%

In analogy to charge transport in metals \cite{AMermin} and magnon transport \cite{magnonWF,ReviewMagnon} in FMs,
we refer to $  {\cal{S}}_{\sigma } \equiv  L_{12\sigma }/(T  L_{11\sigma }) $ as the antiferromagnetic magnon Seebeck coefficient  
and $ {\cal{P}}_{\sigma } \equiv  L_{21\sigma }/L_{11\sigma } $ as the antiferromagnetic Peltier coefficient for up and down magnons.
The Onsager relation provides the Thomson relation (also known as Kelvin-Onsager relation \cite{spincal}) ${\cal{P}}_{\sigma } =  T  {\cal{S}}_{\sigma }$.
In contrast to FMs, the total term vanishes for AFs,
$  {\cal{S}}_{\rm{AF}} \equiv \sum_{\sigma } {\cal{S}}_{\sigma } =0 $ and
$  {\cal{P}}_{\rm{AF}}  \equiv  \sum_{\sigma } {\cal{P}}_{\sigma } = 0 $,
due to the opposite magnetic dipole moment, $ \sum_{\sigma }  {{ L}}_{12\sigma }  =  \sum_{\sigma }  {{ L}}_{21\sigma } =0 $.
%%%%%%%%%%%%%%%%%%%%%%%%%%%%
Still, focusing on each magnon mode separately, the coefficients ${\cal{S}}_{\sigma }$ and $ {\cal{P}}_{\sigma }$ show a universal behavior at low temperature, $b \equiv   \Delta/(k_{\rm{B}} T) \gg  1$, in the sense that the coefficients do not depend on the antiferromagnetic exchange interaction $J$ specific to the material,
and reduce to the qualitatively same form as the ones for FMs, \cite{magnonWF,ReviewMagnon}
\begin{eqnarray}
 {\cal{S}}_{\sigma } \stackrel{\rightarrow }{=}    \frac{\sigma }{g \mu _{\rm{B}}} \frac{\Delta}{T}, \,\,\,\,\,\,\,\,\,\,\,\,\,
 {\cal{P}}_{\sigma }  \stackrel{\rightarrow }{=}  \sigma \frac{\Delta}{g \mu _{\rm{B}}}.
\label{eqn:Peltier}  
\end{eqnarray}
This is another demonstration of the fact that the dynamics of the AF reduces to independent copies for each magnon $\sigma =\pm 1$ in FMs; 
up and down magnons are completely decoupled in an AF in leading magnon-approximation given by Eq. (\ref{eqn:H2}).
%%%%%%%%%%%%%%%%%%%%%%%%%%%%%
This implies that the up and down contributions,  $K_{\sigma }$, to the thermal conductance of the AF $K_{\rm{AF}} = \sum_{\sigma } K_{\sigma } $ can be considered separately, and are given by \cite{magnonWF,ReviewMagnon}
\begin{eqnarray}
K_{\sigma } = \frac{1}{T} \Big({{ L}}_{22\sigma } - \frac{{{ L}}_{12\sigma }{{ L}}_{21\sigma }}{{{ L}}_{11\sigma }} \Big).
   \label{eqn:Kaf}
\end{eqnarray}
As we have seen in the study of FMs \cite{magnonWF,KJD,ReviewMagnon}, $K_{\sigma }$ is expressed by off-diagonal elements \cite{GrunwaldHajdu,KaravolasTriberis} ${{{ L}}_{12\sigma }{{ L}}_{21\sigma }}/{{{ L}}_{11\sigma }}$ as well as ${{ L}}_{22\sigma }$.
This can be seen in the following way (see Refs. [\onlinecite{magnonWF,KJD,ReviewMagnon}] for details).
The applied temperature gradient $ \partial _x  T$ induces a magnonic spin current for each magnon ($\sigma  = \uparrow , \downarrow $), 
$ \langle j_{x \sigma }  \rangle =  - L_{12\sigma }\partial _x  T/T$, which leads to an accumulation of each magnon at the boundaries and thereby builds up a non-uniform magnetization since two magnon modes are decoupled and do not interfere with each other in the AF.
This generates an intrinsic magnetization gradient \cite{SilsbeeMagnetization,Basso,Basso2,Basso3,MagnonChemicalWees,YacobyChemical} $\partial _x  B_{\sigma }^{\ast }$ acting separately on each magnon that produces a magnonic counter-current.
Then, the system reaches a stationary state such that in- and out-flowing magnonic spin currents balance each other;  $\langle j_{x \sigma }  \rangle = 0$ in this new quasi-equilibrium state where 
\begin{eqnarray}
   \partial _x  B_{\sigma }^{\ast } =  \frac{L_{12\sigma }}{L_{11\sigma }}   \frac{ \partial _x  T}{T}.
  \label{eqn:eachB}
\end{eqnarray}
Thus the total thermal conductance $K_{\rm{AF}}= \sum_{\sigma } K_{\sigma } $ defined by $  \langle j_{x\sigma }^Q \rangle = - K_{\sigma } \partial _x  T$ is measured.
Since the thermally-induced intrinsic magnetization gradient $\partial _x  B_{\sigma }^{\ast }$ given in Eq. (\ref{eqn:eachB}) acts individually on each magnon as an effective magnetic field gradient, inserting Eq. (\ref{eqn:eachB}) into Eq. (\ref{eqn:2by2af}), the contribution from each magnon $ K_{\sigma }$ to the total thermal conductance of the AF $K_{\rm{AF}} $ becomes Eq. (\ref{eqn:Kaf}) in terms of Onsager coefficients where the off-diagonal elements \cite{GrunwaldHajdu,KaravolasTriberis} arise from the magnetization gradient-induced counter-current. 
This is in analogy to  thermal transport of electrons in metals \cite{AMermin} where, however, the off-diagonal contributions are  strongly suppressed by  the sharp Fermi surface of fermions at  temperatures $k_BT$ much smaller than  the Fermi energy.

The coefficient $L_{11\sigma }$ is identified with the magnonic spin conductance $ G_{\sigma }= L_{11\sigma }   $ for each magnon and the total one of the AF is given by $ G_{\rm{AF}} =  \sum_{\sigma } G_{\sigma } $. From these we obtain the thermomagnetic ratio $K_{\rm{AF}}/G_{\rm{AF}}  $, characterizing magnonic spin and thermal transport in the AF. At low temperatures, $  b\gg 1  $, the ratio  becomes  linear in temperature,
\begin{eqnarray}
 \frac{K_{\rm{AF}}}{G_{\rm{AF}}} \stackrel{\rightarrow }{=} 
 {\cal{L}}_{\rm{AF}} T \, .
   \label{eqn:WFaf}
\end{eqnarray}
Here, ${\cal{L}}_{\rm{AF}}$ is the magnetic Lorenz number for AFs given by
\begin{eqnarray}
 {\cal{L}}_{\rm{AF}} =   \frac{5}{2} \Big(\frac{k_{\rm{B}}}{g \mu _{\rm{B}}}\Big)^2,
   \label{eqn:LorentzAF}
\end{eqnarray}
which is independent of material parameters apart from the $g$-factor.
Thus, at low temperatures, the ratio $ {K_{\rm{AF}}}/{G_{\rm{AF}}}$ satisfies the WF law in the sense that it becomes linear in temperature; the WF law holds in the same way for magnons both in AFs and FMs \cite{magnonWF,ReviewMagnon,KJD}, which are bosonic excitations, as for electrons \cite{WFgermany,AMermin} which are fermions. In this sense, it can be concluded that the linear-in-$T$ behavior of the thermomagnetic ratio is indeed universal.
We remark that if one wrongly omits the off-diagonal coefficients \cite{GrunwaldHajdu,KaravolasTriberis} in Eq. (\ref{eqn:Kaf}) which can be as large as the diagonal ones, the ratio would not obey WF law, breaking the linearity in temperature.

Lastly we comment on the factor `$5/2$' in Eq. (\ref{eqn:LorentzAF}) which is different from `1' that we have derived in our last work on magnon transport in topologically trivial three-dimensional ferromagnetic junctions.~\cite{magnonWF,ReviewMagnon}
The difference arises from the geometry of the system setup, single bulk or junction, rather than FMs or AFs. Indeed, the factor `$5/2$' arises also in a single bulk FM and the ratio reduces to the same form Eq. (\ref{eqn:WFaf}). This can be seen also as follows; until now, considering a topologically trivial three-dimensional single bulk AF, we have seen that the up and down magnons of the AF  are completely decoupled (in leading order) and the dynamics indeed reduces to independent copies of each magnon in single bulk FMs [Eq. (\ref{eqn:H2})].
Therefore focusing only on $\sigma$-magnons and using Eqs. (\ref{eqn:11})-(\ref{eqn:22}), the magnonic WF law for a single bulk FM can be derived, which becomes at low temperatures, $ b\gg 1  $,
\begin{eqnarray}
 \frac{K_{\sigma  }}{G_{\sigma  }} \stackrel{\rightarrow }{=} \frac{5}{2}  \Big(\frac{k_{\rm{B}}}{g \mu _{\rm{B}}}\Big)^2    T.
   \label{eqn:eachWFaf}
\end{eqnarray}
Thus, we see that the factor `$5/2$' arises also for the single bulk FM \footnote{This well agrees with a recent calculation by A. Mook {\it{et al.}} \cite{MookPrivate} of the magnonic WF law for a single bulk ferromagnetic insulator.}, and we conclude that the factor `$5/2$' is common to both ferro- and antiferromagnetic single bulk magnets.
From this, we see that in contrast to the universality of the linear-in-$T$ behavior, the magnetic Lorenz number is not; it may vary from system to system depending on {\it{e.g.}}, the geometry of the setup, single bulk or junction \cite{magnonWF}, and the system dimension \cite{KJD}.
%%%%%%%%%%%%%%%%%%%%%%%%%
In addition, the Onsager coefficients for the systems depend on the details of the setup by having different types of polylogarithm function ${\rm{Li}}_{s}$; both ferro- and antiferromagnetic single bulk magnets are described by ${\rm{Li}}_{l+3/2}$ ($l = 0, 1, 2$), while the three-dimensional ferromagnetic junction  by  the exponential integral ${\rm{Li}}_{l}$. This difference in the polylogarithm functions gives rise to different prefactors depending on the system setup.

%%%%%%%%%%%%%%%
\section{Topological AF}
\label{sec:nontrivial}
%%%%%%%%%%%%%%%

In this section, using above results, we consider a clean AF on a two-dimensional square lattice ($d=2$), embedded in the $xy$-plane, with a focus on the effects of an electric field ${\mathbf{E}}$ that couples to the magnetic dipole moment $ \sigma  g \mu _{\rm{B}}{\mathbf{e}}_z$ of up and down magnons  through the AC effect \cite{casher}.

%%%%%%%%%%%%%%%%%%%%%%%%%%%%%%
\subsection{Aharonov-Casher effect on magnons}
\label{subsec:ACmagnon}
%%%%%%%%%%%%%%%%%%%%%%%%%%%%%%

In the last section (Sec. \ref{sec:trivial}), starting from the spin Hamiltonian Eq. (\ref{eqn:H}) in the absence of electric fields, we have shown that the low-energy dynamics of AFs in the long wave-length (continuum) limit is described by the completely decoupled up and down magnons, see Eq. (\ref{eqn:H2}), with dispersion $\hbar \omega _{{\mathbf{k}}\sigma } = Dk^2 + \Delta_{\sigma }$.
We can then introduce an effective Hamiltonian for such magnon modes, given by ${\cal{H}}_{\rm{m} \sigma }  =  {\hat{{\mathbf{p}}}}^2/2m  + \Delta_{\sigma }$, where the effective mass of the magnons is defined by $(2m)^{-1} = D/\hbar ^2$ with $d=2$, and $\hat{{\mathbf{p}}} =(p_x, p_y, 0)$ is the momentum operator.  
Thus, the magnons behave like ordinary particles of mass $m$ with quadratic dispersion, moving in the $xy$-plane and carrying a magnetic dipole moment $ \sigma  g \mu _{\rm{B}}{\mathbf{e}}_z$.

In the presence of an electric field ${\mathbf{E}}({\mathbf{r}})$, the magnetic dipole moment $ \sigma  g \mu _{\rm{B}}{\mathbf{e}}_z$ of a moving magnon experiences a magnetic force in the rest frame of the magnon. This system is formally identical to the one  studied by Aharonov and Casher~\cite{casher}, namely that 
of a neutral particle carrying a magnetic dipole moment, moving in an electric field.
Thus, following their work~\cite{casher} we account for the electric field by replacing the momentum operator $ \hat{{\mathbf{p}}}  $  by $\hat{{\mathbf{p}}} +  \sigma  {g \mu _{\rm{B}}} {\mathbf{A}}_{\rm{m}}/{c}$, where
\begin{eqnarray}
{\mathbf{A}}_{\rm{m}}({\mathbf{r}})
 = \frac{1}{c} {\mathbf{E}}({\mathbf{r}})\times {\mathbf{e_z}}
\label{magnon_gauge}
\end{eqnarray}
is the `electric' vector potential  ${\mathbf{A}}_{\rm{m}}$ acting on the magnons
at position ${\mathbf{r}}=(x,y,0)$.
The total Hamiltonian then becomes $  {\cal{H}}_{\rm{m}} = \sum_{\sigma=\pm }  {\cal{H}}_{\rm{m} \sigma }$ with 
\begin{eqnarray}
 {\cal{H}}_{\rm{m} \sigma }  =    \frac{1}{2m} \Big(\hat{{\mathbf{p}}} +  \sigma  \frac{g \mu _{\rm{B}}}{c}{\mathbf{A}}_{\rm{m}} \Big)^2 + \Delta_{\sigma }.
\label{HamiltonianLL} 
\end{eqnarray}
This expression  describes the low-energy dynamics of the magnons moving in an electric field ${\mathbf{E}}$.
It is valid at sufficiently low temperatures where the effects of magnon-magnon and magnon-phonon interactions become \cite{magnonWF,adachiphonon,Tmagnonphonon} negligibly small.
The Hamiltonian in Eq.~(\ref{HamiltonianLL}) is formally identical to that of a charged particle moving in a magnetic vector potential, in which the coupling constant is given by $\sigma g \mu _{\rm{B}}$ instead of the electric charge $e$. 

When an electric field has the special quadratic form ${\mathbf{E}}({\mathbf{r}}) = {\mathcal{E}}(-x/2, -y/2,0)$, with $ {\mathcal{E}}$ a constant field gradient,
it gives rise to the  `symmetric' gauge potential
${\mathbf{A}}_{\rm{m}}({\mathbf{r}}) = ({\mathcal{E}}/c)(-y/2, +x/2,0)$.  
Since 
\begin{eqnarray}
  {\mathbf{\nabla }} \times  {\mathbf{A}}_{\rm{m}} = \frac{{\mathcal{E}}}{c} {\mathbf{e}}_z,
  \label{eqn:correspondenceEB} 
\end{eqnarray}
the field gradient $ {\mathcal{E}}$ plays the role of the perpendicular magnetic field in two-dimensional electron gases \cite{Ezawa}. The considered electric field with the constant gradient $\cal{E}$ can be realized $\it{e.g.}$ by an STM tip \cite{GeimSTM,STM_Egradient}.
Similarly, the analog of the Landau gauge ${\mathbf{A}}_{\rm{m}}({\mathbf{r}}) = ({\mathcal{E}}/c)(0, x,0)$ is provided by 
${\mathbf{E}}({\mathbf{r}})= {\mathcal{E}}(-x, 0,0)$.  
Within the quantum-mechanical treatment \cite{Ezawa}, the resulting magnon dynamics [Eqs. (\ref{LL})-(\ref{CM})] is identical to the one of the symmetric gauge since both satisfy $ {\mathbf{\nabla }} \times  {\mathbf{A}}_{\rm{m}} = ({{\mathcal{E}}}/{c}) {\mathbf{e}}_z$ and thus the two Hamiltonians with different `gauges' can be transformed into each other by the unitary gauge transformation $U_\sigma \equiv  {\rm{exp}}(i \sigma  g \mu _{\rm{B}} {\cal{E}} xy/2 \hbar  c^2)$.
%%%%%%%%%%%%%%

With the Hamiltonian ${\cal{H}}_{\rm{m} \sigma }$ given in Eq.~(\ref{HamiltonianLL}) we can adopt the topological formulations \cite{NiuBerry,Kohmoto} of the conventional QHE in terms of Chern numbers. See Ref. [\onlinecite{KJD}] for the developed formulation of AC phase-induced magnon Hall effects in FM [see also Fig. \ref{fig:HelicalAFChiralFM} (a)], which corresponds to that for $ {\cal{H}}_{\rm{m} \uparrow  }$.

Using canonical equations, $ \dot{\mathbf{r}}= {\mathbf{v}} =   \partial {\cal{H}}_{\rm{m} \sigma }/\partial {\mathbf{p}} $ and $ \dot{\mathbf{p}} = -  \partial {\cal{H}}_{\rm{m} \sigma }/\partial {\mathbf{r}}$, where $\dot{\mathbf{r}} $ denotes the time derivative of $\mathbf{r}$ and ${\mathbf{v}}$ is the velocity, the force ${\mathbf{F}}_{\rm{AC}}$ acting on magnons in electric fields is then given by~\cite{magnon2}
\begin{eqnarray}
{\mathbf{F}}_{\rm{AC}} = \sigma  g \mu_{\rm{B}} \Big[{\mathbf{\nabla }}B - \frac{{\mathbf{v}}}{c} \times ({\mathbf{\nabla }} \times  {\mathbf{A}}_{\rm{m}})\Big]. 
\label{ForceAC}
\end{eqnarray}
The force ${\mathbf{F}}_{\rm{AC}} $ is invariant under the gauge transformation 
$ {\mathbf{A}}_{\rm{m}} \mapsto {\mathbf{A}}_{\rm{m}}'=  {\mathbf{A}}_{\rm{m}} +  {\mathbf{\nabla }}\chi  $ accompanied by ${\mathbf{E}} \mapsto {\mathbf{E}}' = {\mathbf{E}} + c {\mathbf{e}}_z \times {\mathbf{\nabla }}\chi  $ for arbitrary  scalar function $\chi  = \chi (x,y)$.
Note that the gauge invariance in the present case is specific  to electrically neutral particles only, such as magnons, since $ {\mathbf{E}}'$ and ${\mathbf{E}}$ give rise to different physical forces on charged particles. 
Inserting Eq. (\ref{eqn:correspondenceEB}) into Eq. (\ref{ForceAC}), the force becomes
\begin{eqnarray}
{\mathbf{F}}_{\rm{AC}} = m \dot{\mathbf{v}} = \sigma  g \mu_{\rm{B}} \Big({\mathbf{\nabla }}B - {\mathbf{v}} \times \frac{{\mathcal{E}} {\mathbf{e}}_z}{c^2} \Big),
\label{ForceAC2}
\end{eqnarray}
which indicates that the role of electric field and magnetic field in electrically charged particles \cite{Ezawa}  is played by the magnetic field gradient ${\mathbf{\nabla }}B$ and the electric field gradient ${\mathcal{E}}$, respectively, for `magnetically charged' particles such as our magnons.
%%%%%%%%%

Assuming that the velocity ${\mathbf{v}}$ consists of the cyclotron motion ${\mathbf{v}}_{\rm{c}}$ (see Sec. \ref{subsec:MQSHE}) 
and the drift velocity ${\mathbf{v}}_{\rm{d}} $ with \cite{Ezawa} ${\mathbf{\dot{v}}} _{\rm{d}}=0$, Eq. (\ref{ForceAC2}) gives $ {\mathbf{v}}_{\rm{d}} \times {\mathbf{e}}_z = (c^2/{\cal{E}}) {\mathbf{\nabla }}B  $.
Applying the magnetic field gradient along the $x$ axis $\partial _x B \not=0$ while $\partial _y B =0$, the drift velocity becomes 
${\mathbf{v}}_{\rm{d}}=(0, c^2 \partial _x B/{\cal{E}}, 0) $, 
which is perpendicular to the applied magnetic field gradient and independent of $\sigma$. 
Thus, in the presence of both magnetic and electric field gradients, each magnon ($\sigma = \pm 1$) performs the drift motion along the same direction since both driving forces depend on $\sigma$ [see Eq. (\ref{ForceAC2})] and eventually the $\sigma$-dependence cancels out as $ \sigma^2 =1  $.
This is consistent with the results of `bulk' Hall conductances given in Eq. (\ref{eqn:ParticleHall}) where (Sec. \ref{subsec:TopoHallBulk}) the magnetic field gradient is applied perturbatively. The matrix element in Eq. (\ref{eqn:2by22222}) for magnonic spin Hall effects of bulk magnons [Eq. (\ref{eqn:G_magnonHall77})] indeed vanishes due to the relation between each topological integer. See Eqs. (\ref{eqn:GKzero})-(\ref{eqn:ParticleHall}) for details.

Note that the drift velocity vanishes in the absence of the magnetic field gradient, while each magnon ($\sigma = \pm 1$) still performs the cyclotron motion in opposite directions due to the electric field gradient [Eqs. (\ref{ForceAC2})-(\ref{CM})], leading to the helical edge magnon states, and we consider this situation henceforth.
See Sec. \ref{subsec:MQSHE} (also Appendix \ref{sec:CyclotronMotion}) for details.

%%%%%%%%%%%%%%%%%%%%%%%%%%%%%%
\subsection{A bosonic analog of QSHE by edge magnons}
\label{subsec:MQSHE}
%%%%%%%%%%%%%%%%%%%%%%%%%%%%%%

A straightforward calculation using Eq. (\ref{HamiltonianLL}) with $B=0$ (see Appendix \ref{sec:CyclotronMotion} for details) shows that the quantum dynamics \cite{Ezawa} of down and up magnons are identical  except that the direction of their cyclotron motion is opposite (Fig. \ref{fig:HelicalAFChiralFM}). Indeed, they form the same Landau levels \cite{KJD} with the principal quantum number $n_{\sigma }\in  {\mathbb{N}}_0$,
\begin{eqnarray}
E_{n_{\sigma }}   = \hbar \omega _c \Big(n_{\sigma } +  \frac{1}{2}\Big)  + \Delta   \    \    \      {\rm{for}}     \     \    n_{\sigma } \in  {\mathbb{N}}_0,
\label{LL} 
\end{eqnarray}
and the two magnons  $ \sigma  g \mu _{\rm{B}}{\mathbf{e}}_z$ perform cyclotron motions with the same frequency \cite{KJD}
\begin{eqnarray}
\omega _c  =  \frac{g \mu _{\rm{B}}{\mathcal{E}}}{m c^2}
\label{omega_c} 
\end{eqnarray}
and same electric length \cite{KJD} $ l_{\rm{{\mathcal{E}}}}$,  defined by
\begin{eqnarray}
 l_{\rm{{\mathcal{E}}}} \equiv  \sqrt{{\hbar c^2}/{g \mu _{\rm{B}}{\mathcal{E}}}},
\label{l_E} 
\end{eqnarray}
but along opposite direction, cf. Fig. \ref{fig:HelicalAFChiralFM} (b),
\begin{eqnarray}
  \frac{d}{dt}({\cal{R}}_{x \sigma } + i {\cal{R}}_{y \sigma })=  i \sigma   \omega _c  ({\cal{R}}_{x \sigma } + i {\cal{R}}_{y \sigma }),
  \label{CM} 
\end{eqnarray}
where  ${\mathbf{R}}_{{\rm{{\mathcal{E}}}}\sigma } =({\cal{R}}_{x \sigma }, {\cal{R}}_{y \sigma }) $  is the relative coordinate \cite{SeeAppendix}.
The factor $\sigma$ in Eq. (\ref{CM}) is rooted in the magnetic dipole moment $ \sigma  g \mu _{\rm{B}}{\mathbf{e}}_z$ of a magnon.
The source of cyclotron motion is the electric field gradient $\cal{E}$ [Eqs. (\ref{eqn:correspondenceEB}) and (\ref{omega_c})], which is common to the both modes.

\begin{figure*}[t]
\begin{center}
\includegraphics[width=18cm,clip]{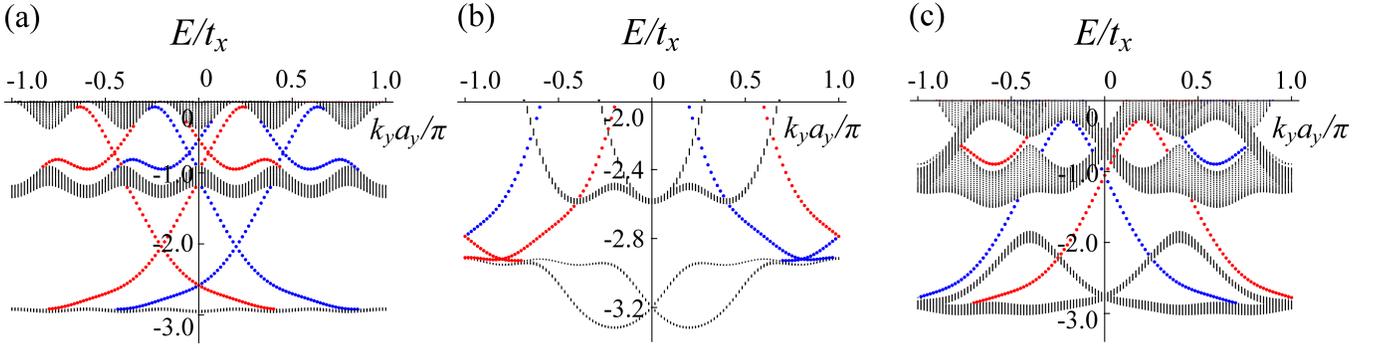}
\caption{(Color online)
Plots of the magnonic energy spectrum, rescaled energy $E/t_x$, as function of the rescaled wavevector $k_y a_y/\pi$ obtained by numerically solving the tight-binding model Eq. (\ref{TightBinding2}) for a strip with width of 115 lattice sites, for $t_x=t_y = 1$ and $\theta _1=2\pi/5$, showing the first and partially the second Landau levels (black). The up magnon edge states ($\sigma =1$) are in red while down magnon edge states ($\sigma =-1$) are in blue.
The periodicity of the vector potential is (a) $ q \gg  1$, (b) $q = 7$, and (c) $q = 4$.
Panel (a) shows a band structure characteristic for a TI with a well-developed gap in an almost flat band and with helical edge  states formed by the up and down magnons.
Panel (b): Qualitatively the same as in (a).
Panel (c): For each given value of $k_y$ there are well-defined edge states with a gap in the spectrum, but they coexist with bulk states at different momenta. 
}
\label{fig:HelicalEdge}
\end{center}
\end{figure*}

Taking into account the opposite directions of the cyclotron motion, the quantum dynamics of the AF reduces to independent copies \cite{Z2topo,TIreview,TIreview2} for each magnon $\sigma =\pm 1$ in FMs (Fig. \ref{fig:HelicalAFChiralFM}) $  {\cal{H}}_{\rm{m}} = \sum_{\sigma }  {\cal{H}}_{\rm{m} \sigma }$.
In Ref. [\onlinecite{KJD}], we have shown that at low temperature $ k_{\rm{B}} T \ll    \hbar   \omega _c $, only the lowest energy mode $n_{\uparrow }=0$ in Eq. (\ref{LL}) becomes relevant and the cyclotron motion of up magnons along one direction [Fig. \ref{fig:HelicalAFChiralFM} (a)] leads to a {\it{chiral}} edge state giving \cite{HalperinEdge,BulkEdgeHatsugai,RShindou,RShindou2,RShindou3} the Chern number \footnote{See Ref. [\onlinecite{KJD}] for the definition of the Berry curvature which gives the Chern number.} $ {\cal{N}}_{0 \uparrow } = +1$.
Since the dynamics of down magnons is the same as that of up magnons except that the direction of cyclotron motion is opposite [Fig. \ref{fig:HelicalAFChiralFM} (b)], the down magnon propagates also along the edge of the sample but in the opposite direction to that of the up magnon, which gives \cite{HalperinEdge,BulkEdgeHatsugai,RShindou,RShindou2,RShindou3} the Chern number ${\cal{N}}_{0 \downarrow } = -1$.
%%%%%%%%%%%%%%%%%%%%%%%%%%%%%%%%%%%%%%%%
Thus, at low temperatures,  $ k_{\rm{B}} T \ll    \hbar   \omega _c $, the Chern number ${\cal{N}}_{0 \sigma }$ of up and down magnons in the lowest Landau level $n_{\sigma } = 0$ is summarized by 
\begin{eqnarray}
{\cal{N}}_{0 \sigma }  = \sigma  ,
\label{eqn:eachChernNumber} 
\end{eqnarray}
and the AF is characterized by the resulting {\it{helical}} edge magnon state (Fig. \ref{fig:HelicalEdge}) where due to the opposite magnon spin $\sigma =\pm 1$, up and down magnons propagate along the edge of the sample but in opposite direction \cite{QSHE2016} [Fig. \ref{fig:HelicalAFChiralFM} (b)].
This is a bosonic analog of the QSHE \cite{QSHE2005,QSHE2006,Z2topo,Z2topoHaldane,Z2SM,TIreview,TIreview2} for electronic edge states, namely, the QSHE for edge magnons induced by the AC effect. 
%%%%%%%%%%%%%%%%%%%
Note that due to the opposite cyclotron motion, the total Chern number vanishes,
\begin{eqnarray}
\sum_{\sigma } {\cal{N}}_{0 \sigma }  =  {\cal{N}}_{0 \uparrow }+ {\cal{N}}_{0 \downarrow } = 0,
\label{eqn:ChernPlus} 
\end{eqnarray}
while
\begin{eqnarray}
%{\cal{Z}}_{0} \equiv \frac{{\cal{N}}_{0 \uparrow } - {\cal{N}}_{0 \downarrow }}{2}   =   1     \   \   \  ({\rm{mod}} \,\, 2).
{\cal{Z}}_{0} \equiv \frac{1}{2}({\cal{N}}_{0 \uparrow } - {\cal{N}}_{0 \downarrow } )  =   1     \   \   \  ({\rm{mod}} \,\, 2).
\label{eqn:Z2} 
\end{eqnarray}
Thus the QSHE of helical edge magnon states is characterized by a ${\mathbb{Z}}_{2}$ topological number\cite{Z2topo,Z2topoHaldane,Z2SM,TIreview,TIreview2,Z2Robust}, explicitly given here by  ${\cal{Z}}_{0} = 1$, and the AF with the AC effect may be identified \cite{ETI,ETI2,PhotonTopo,PhotonTopo3D} with a bosonic version \cite{QSHE2016} of TIs. Such a magnonic analog of TIs can be understood as copies \cite{Z2topo,TIreview,TIreview2} of the ferromagnetic `quantum' Hall system \cite{KJD} having opposite magnon polarization.

%%%%%%%%%%%%%%%%%%%%%%%%%%%%%%
\subsection{Energy spectrum and chiral edge states}
\label{subsec:LatticeNumCal}
%%%%%%%%%%%%%%%%%%%%%%%%%%%%%%

We calculate the magnon energy spectrum now  for a finite geometry of strip shape to find the Landau levels and in particular the chiral edge modes of magnons, all in analogy to the QHE for electrons~\cite{JKDL2013,JKDL2014}. For the numerical evaluation we need to discretize the continuum Hamiltonian  $ \sum_{\sigma}{\cal{H}}_{\rm{m} \sigma } $, given in Eq. (\ref{HamiltonianLL}). This leads to the standard TBR of a continuum Hamiltonian in the presence of a gauge potential~\cite{Textbook_TightBindingRep,TBrepDatta,TightBindingRep,TBrepQHE},
\begin{eqnarray}
 {\cal{H}}_{\rm{AC}}  =   -  \sum_{\sigma = \uparrow , \downarrow } \sum_{\langle i j \rangle} (t_{ij} {\rm{e}}^{ i \sigma  \theta _{ij}}
a_{i, \sigma } a_{j, \sigma }^\dagger + {\rm{H.c.}}),
 \label{magnon_hopping}  
\end{eqnarray}
where 
$a_{i \sigma}$ is the annihilation operator of spin-$\sigma$ magnons localized at the site $i$ satisfying the bosonic commutation relations, $[a_{i \sigma }, a_{j \sigma^{\prime} }^\dagger] = \delta _{i j} \delta _{\sigma \sigma^{\prime}}  $ {\it{etc.}}, and where the Peierls phase
$\theta _{ij} = (g \mu _{\rm{B}}/\hbar c) \int_{{\mathbf{r}}_i}^{{\mathbf{r}}_j} d  {\mathbf{r}} \cdot  {\mathbf{A}}_{\rm{m}}   $ 
is the AC phase which the magnon with the magnetic moment $ \sigma  g \mu _{\rm{B}}{\mathbf{e}}_z$  acquires during the hopping on the lattice, and $t_{ij} >0 $ is the hopping amplitude. 
Here, we suppressed the  constant ${\Delta}$ [Eq. (\ref{HamiltonianLL})], being irrelevant for the chiral edge states. 
If a magnon hops between site $i$ and $j$ along $x\, (y)$ direction, the amplitude is given by \cite{TBrepDatta} $t_{x(y)}  = \hbar ^2/(2m a_{x(y)}^2) $, where $a_{x(y)}$ is the lattice constant along $x (y)$ direction in the TBR.
For simplicity, we will consider the isotropic limit $t_x=t_y$ henceforth.
In the continuum limit, $a_{x,y}\to 0$, 
Eq. (\ref{magnon_hopping}) reduces to the magnon Hamiltonian Eq. (\ref{HamiltonianLL}).

We wish to emphasize that the tight-binding lattice is just introduced
for calculational purposes and the tight-binding lattice is not related to the original lattice of the spin system, Eq.~(\ref{eqn:H}), from which we started. In other words, there is no relation between the lattice
constants $a_{x(y)}$ occurring in the TBR and the lattice constants occurring in Eq.~(\ref{eqn:H}). Also, searching  for edge states that are topological and thus independent of microscopic details, we can choose parameter values in the simulations  that are most convenient from a  numerical point of view.

%\blue{This is in contrast to the work \cite{KJD} on FMs.}

%To perform exact numerical diagonalization of the tight-binding Hamiltonian Eq. (\ref{magnon_hopping}), 
Next, we use the analog of the Landau gauge such that the system is translation-invariant along the $y$ axis. Introducing the momentum $k_y$, we can perform a Fourier transformation of Eq. (\ref{magnon_hopping}) such that ${\cal{H}}_{\rm{AC}} =\sum_{k_y} H_{k_y}$,\cite{JKDL2013,JKDL2014}, with
\begin{align}
H_{k_y}&=-t_x \sum_{n,\sigma} (a_{k_y,n+1,\sigma}^\dagger a_{k_y,n,\sigma} + {\rm{H.c.}} ) \label{TightBinding2}   \\
            &\hspace{11pt}- 2 t_y  \sum_{n,\sigma}  [ \cos (k_y a_y+\sigma \theta_n)] a_{k_y,n,\sigma}^\dagger a_{k_y,n,\sigma},         \nonumber 
\end{align}
where $a_{k_y,n,\sigma}$ annihilates a spin-$\sigma $ magnon with momentum $k_y$ in $y$ direction at site $n=x/a_x$ (along $x$ direction). 
The AC phase accumulated by the up magnon ($\sigma =1$) as it hops in  $y$ direction by one lattice constant $a_y$ is given by  
$\theta _{n} =(g \mu _{\rm{B}}/\hbar c^2) \mathcal{E} n a_x a_y = n \theta _1 $, 
where $\theta_1 \equiv (g \mu _{\rm{B}}/\hbar c^2) \mathcal{E} a_x a_y$, while the down magnon ($\sigma =-1$) acquires the opposite sign $ - \theta _{n}$.
For definiteness, we focus on the spectrum around the lowest Landau level.

Performing exact numerical diagonalization of the Hamiltonian~(\ref{TightBinding2}), we obtain the spectrum shown in Fig. \ref{fig:HelicalEdge}.
In the TI regime, the system hosts a pair of helical edge magnon states. We have checked numerically, that choosing different parameter values changes the spectrum quantitatively but the helical edge states remain, showing that they are indeed topologically stable.

To avoid a breakdown of the sample due to the huge voltage drop resulting from an applied strong electric field, we also consider  electric fields and vector potentials ${\mathbf{A}}_{\rm{m}}$ that are periodic in $x$ direction and of saw-tooth shape~\cite{KJD}.  Using such a periodically extended fields only over a distance that can be much smaller than the sample dimensions or even the electric length $ l_{\rm{{\mathcal{E}}}}$, we \cite{KJD} have seen that the requirement of strong field gradients ${\mathcal{E}}$ needed for creating a quantum Hall effect of magnons in FMs ({\it{e.g.}}, Landau levels and the resultant level spacing) can be substantially softened, while still producing well-defined chiral edge magnon states \cite{KJD}, since the magnitude of ${\mathcal{E}}$ for each period remains the same. 
Periodic fields may be realized by periodically arranging STM tips \cite{GeimSTM,STM_Egradient,KJD}.
%%%%%%%%%%%%%%%%%%%%

Such periodic potentials are easily implemented in our approach by assuming in Eq. (\ref{TightBinding2}) $\theta_{n} = \theta_1 q \{n/q\}$, with period $q$ (integer) and where $\{\cdot \}$ denotes the fractional part smaller than one. This implies that  the periodic vector potential has the form $ {\mathbf{A}}_{{\rm{m}}q} = ({\mathcal{E}} R_q/c) (0, \{x/R_q \}, 0) $, where  $R_q= q a_x$ is the period.

From Fig. \ref{fig:HelicalEdge} we see that for large period $q$ there is a well-developed gap in an almost flat band and with the corresponding edge states, see Fig. \ref{fig:HelicalEdge} (a). If $q$ gets smaller than the electric length, the bulk gap is no longer uniform in momentum, see Figs. \ref{fig:HelicalEdge} (b) and (c). 
As a result, edge states coexist with bulk modes at different momenta \cite{Li2016,MagnonicWeylSemimetalAC}, see Fig. \ref{fig:HelicalEdge} (c). 
However, for fixed values of $k_y$, there is still a gap in the spectrum and furthermore, well-defined edge states still exist.
Thus, if disorder is weak the edge modes will not couple to the bulk and the Hall conductance will still be dominated by these edge modes, similarly to Weyl semimetals.

Under the assumption that the spin along the $z$ direction remains a good quantum number \cite{Z2Robust,JelenaYTwire}, we have seen that the key to a nonzero Chern number $ {\cal{N}}_{0 \sigma } $ is the cyclotron motion of individual magnons. Indeed, the AC phase-induced cyclotron motion leads to edge magnon states (Fig. \ref{fig:HelicalEdge}) each giving \cite{HalperinEdge,BulkEdgeHatsugai,RShindou,RShindou2,RShindou3} rise to a non-zero  Chern number $ {\cal{N}}_{0 \sigma }  = \sigma   $. Therefore, as long as magnons can perform cyclotron motions, the edge magnon state is robust against external perturbation \cite{QSHE2005,QSHE2006,Z2topo,Z2topoHaldane,Z2SM,TIreview,TIreview2} and the relation $  {\cal{N}}_{0 \uparrow }+ {\cal{N}}_{0 \downarrow } = 0  $ between each topological integer \cite{Z2topo,TIreview,TIreview2,Z2Robust} remains valid.
Indeed, it has been confirmed experimentally that magnons satisfy Snell's law at interfaces \cite{Snell_Exp,Snell2magnon}, indicating specular (i.e., elastic) reflection at the boundary to vacuum, and thereby we can expect that magnons form skipping orbits along the boundary like electrons \cite{HalperinEdge}, giving rise to edge states.

We note that there are still general differences \cite{KJD}  to electrons due to the bosonic nature of the magnons. Due to the Bose-distribution function, even in the presence of topological edge states, the Hall transport coefficients of bulk magnons generally cannot be described in terms of a Chern integer. Only in almost flat bands \cite{KJD}, the Hall coefficients become characterized by such a topological invariant that edge magnon states bring about, while  the Hall coefficients are still characterized by the Bose-distribution function (see Sec. \ref{subsec:TopoHallBulk}). This is in contrast to electronic systems.

%%%%%%%%%%%%%%%%%%%%%%%%%%
\subsection{Hall conductances of magnons}
\label{subsec:TopoHallBulk}
%%%%%%%%%%%%%%%%%%%%%%%%%%

In this section, we discuss Hall transport properties of bulk \cite{HalperinEdge,BulkEdgeHatsugai,RShindou,RShindou2,RShindou3} magnons  in the AC effect-induced magnonic TIs characterized by a spin-dependent Chern number ${\cal{N}}_{0 \sigma } = \sigma $ [Eq. (\ref{eqn:Z2})], by making use of the aforementioned mapping between the system and two independent copies \cite{Z2topo,TIreview,TIreview2} of a ferromagnetic `quantum' Hall system \cite{KJD}. We consider the cases where again the total spin along the $z$ direction is a good quantum number.
%%%%%%%%%%%%%%%%%%%%%%%%%%%%%%%%%%%%%%
The crystal lattice creates a periodic potential for magnons \cite{AMermin,Kohmoto,NiuBerry} $ U({\mathbf{r}}) = U({\mathbf{r}}+ {\mathbf{R}}) $ with Bravais lattice vector ${\mathbf{R}}=(a_x,a_y)$, which gives rise to a band structure for magnons. In the absence of a magnetic field, $B=0$, 
the Hamiltonian for spin-$\sigma$ magnons is given by
$ {\cal{H}}_{\sigma }({\mathbf{r}})   =   {\cal{H}}_{{\rm{m}}\sigma } ({\mathbf{r}})  + U({\mathbf{r}})$.
We then introduce the Bloch Hamiltonian with  Bloch wavevector ${\mathbf{k}}=(k_x, k_y) $ following Refs. [\onlinecite{Kohmoto,NiuBerry,KJD}], $   {\cal{H}}_{{\mathbf{k}}\sigma } \equiv    {\rm{e}}^{- i {\mathbf{k}}\cdot{\mathbf{r}} } {\cal{H}}_{\sigma }  {\rm{e}}^{ i {\mathbf{k}}\cdot{\mathbf{r}} }  = [- i \hbar  {\mathbf{\nabla}} +  \hbar  {\mathbf{k}}  + \sigma  g \mu _{\rm{B}}{\mathbf{A}}_{\rm{m}}({\mathbf{r}})/c]^2/2m + \Delta + U({\mathbf{r}})$, 
where ${\mathbf{A}}_{\rm{m}}$ is the periodically extended vector potential \cite{KJD}.
The eigenfunction of the Schr\"odinger equation ${\cal{H}}_{{\mathbf{k}}\sigma } u_{n {\mathbf{k}}\sigma }({\mathbf{r}}) = E_{n {\mathbf{k}}\sigma } u_{n {\mathbf{k}}\sigma }({\mathbf{r}}) $ is given by \cite{KevinHallEffect,Kohmoto,NiuBerry} the magnonic Bloch wave function 
$ u_{n {\mathbf{k}}\sigma }({\mathbf{r}})  \equiv   {\rm{e}}^{- i {\mathbf{k}}\cdot{\mathbf{r}} }  \psi _{n {\mathbf{k}}\sigma }   $, 
where ${\cal{H}}_{\sigma } \psi _{n {\mathbf{k}}\sigma } = E_{n {\mathbf{k}}\sigma }  \psi _{n {\mathbf{k}}\sigma }$.

At sufficiently low temperature $k_{{\rm{B}}} T  \ll  \hbar \omega _c$, the lowest mode $ n=0$ dominates the dynamics (Fig. \ref{fig:HelicalEdge}).
In Ref. [\onlinecite{KJD}], where we have studied the magnon bands of a FM in the `quantum' Hall phases realized by the electric field gradient-induced AC effects, we have shown that the lowest magnon band is almost flat on the energy scale set by the temperature \cite{KevinHallEffect,RSdisorder,TopoMagBandKagome}, i.e., the band width is much smaller than $  k_{{\rm{B}}} T$, and named it almost flat band.
Due to this flatness, the lowest band [{\it{e.g.}}, Fig. \ref{fig:HelicalEdge} (a)] can be well characterized by its typical energy $E_{0\sigma }^{\ast }$ in the sense that the value of the Bose-distribution function $n_{\rm{B}} (E_{0 {\mathbf{k}}\sigma }) =({\rm{e}}^{\beta E_{0 {\mathbf{k}}}\sigma }-1)^{-1} $ 
with $\beta \equiv (k_{\rm{B}}T)^{-1} $ can be considered as approximately uniform in the Brillouin zone, $n_{\rm{B}} (E_{0 {\mathbf{k}}\sigma }) \simeq n_{\rm{B}} (E_{0 \sigma }^*)$, which we will adopt in the subsequent discussion.

Within the linear response regime, the spin and heat Hall current densities for each mode, $j_{y \sigma } $ and $  j_{y \sigma }^{Q} $, subjected to a magnetic field gradient \cite{Haldane2,ZyuzinRequest} and a temperature one are described by the Onsager matrix
\begin{eqnarray}
%%%%%%%%%%%%%%%%%%%%%%%%%%%%%
\begin{pmatrix}
\langle j_{y \sigma } \rangle    \\  \langle  j_{y \sigma }^{Q}   \rangle
\end{pmatrix}
%%%%%%%%%%%%%%%%%%%%%%%%%%%%%
=
%%%%%%%%%%%%%%%%%%%%%%%%%%%%%
\begin{pmatrix}
 L_{11 \sigma }^{yx}  & L_{12 \sigma }^{yx}    \\    
 L_{21 \sigma }^{yx}  & L_{22 \sigma }^{yx}       
\end{pmatrix}
%%%%%%%%%%%%%%%%%%%%%%%%%%%%%
%%%%%%%%%%%%%%%%%%%%%%%%%%%%%
\begin{pmatrix}
  \partial _x  B    \\  -  \partial _x  T/T
\end{pmatrix}.
%%%%%%%%%%%%%%%%%%%%%%%%%%%%%
\label{eqn:2by2}
\end{eqnarray}
%%%%%%%%%
Since the band is almost flat, the Hall transport coefficients \cite{Matsumoto,Matsumoto2,KJD} $L_{i j \sigma }^{yx}$ can be characterized by the Chern number ${\cal{N}}_{0 \sigma } = \sigma $,
\begin{eqnarray}
L_{i j \sigma }^{yx} =   (k_{\rm{B}}T)^{\eta} (\sigma  g \mu _{\rm{B}})^{2-{\eta}}  
{\cal{C}}_{\eta}  \big(n_{\rm{B}} (E_{0\sigma }^{\ast })\big) \cdot    {\cal{N}}_{0 \sigma }/h,
 \label{eqn:HallLij22}
\end{eqnarray}
where
$ \eta =  i + j -2$, 
${\cal{C}}_0  \big(n_{\rm{B}} (E_{0\sigma }^{\ast })\big)  =   n_{\rm{B}} (E_{0\sigma }^{\ast }) $,
${\cal{C}}_1  \big(n_{\rm{B}} (E_{0\sigma }^{\ast })\big)  =  [1+ n_{\rm{B}} (E_{0\sigma }^{\ast })] {\rm{log}} [1+ n_{\rm{B}} (E_{0\sigma }^{\ast })] 
-  n_{\rm{B}} (E_{0\sigma }^{\ast }) {\rm{log}} [n_{\rm{B}} (E_{0\sigma }^{\ast })] $, and
${\cal{C}}_2  \big(n_{\rm{B}} (E_{0\sigma }^{\ast })\big)  =   [1+ n_{\rm{B}} (E_{0\sigma }^{\ast })]  \big({\rm{log}} [1+ 1/n_{\rm{B}} (E_{0\sigma }^{\ast })]\big)^2 - \big({\rm{log}} [n_{\rm{B}} (E_{0\sigma }^{\ast })]\big)^2 - 2 {\rm{Li}}_2 \big(- n_{\rm{B}} (E_{0\sigma }^{\ast })\big)$.
The Onsager reciprocity is satisfied by having $L_{12 \sigma }^{yx} = L_{21 \sigma }^{yx}$.
%%%%%%%%%%%%%%%
The coefficient $ L_{11 \sigma }^{yx} $ is identified with the magnonic spin Hall conductance $  G_{\sigma}^{yx } $ arising from each magnon and the total one of the AF is given by $ G_{\rm{AF}}^{yx }  =  \sum_{\sigma } G_{\sigma}^{yx } $. The contribution of each magnon $K_{\sigma}^{yx } $ to the thermal Hall conductance of the AF $K_{\rm{AF}}^{yx} = \sum_{\sigma } K_{\sigma}^{yx } $ is expressed in terms of Onsager coefficients by \cite{KJD}
\begin{eqnarray}
    K_{\sigma}^{yx } = \Big( L_{22\sigma }^{yx} - \frac{L_{21\sigma }^{yx} L_{12\sigma }^{yx}}{L_{11\sigma }^{yx}}   \Big)/T,
  \label{eqn:K_magnonHall}
\end{eqnarray}
where as we have seen in Sec. \ref{sec:trivial}, the off-diagonal elements \cite{GrunwaldHajdu,KaravolasTriberis} similarly arise from the magnon counter-current by the thermally-induced magnetization gradient \cite{SilsbeeMagnetization,Basso,Basso2,Basso3,MagnonChemicalWees,YacobyChemical}
$ \partial _x  B_{\sigma }^{\ast } =  ({L_{12\sigma }^{yx}}/{L_{11\sigma }^{yx}})  ({\partial _x  T}/{T})$.
See Ref. [\onlinecite{KJD}] for details of the thermal Hall conductance in the `quantum' Hall regime and the Hall coefficient $L_{i j \uparrow  }^{yx}$.

In the almost flat band $E_{0 \uparrow  }^{\ast }  \approx   E_{0 \downarrow  }^{\ast } \equiv   E_{0 }^{\ast } $, the Hall transport coefficient Eq. (\ref{eqn:HallLij22}) becomes
\begin{eqnarray}
 L_{i j \sigma }^{yx} =    \sigma ^{2-{\eta}}   {L^{\prime}}_{i j}    {\cal{N}}_{0 \sigma },
\label{eqn:spectrum222}
\end{eqnarray}
where we introduced $ {L^{\prime}}_{i j} =   (k_{\rm{B}}T)^{\eta} (g \mu _{\rm{B}})^{2-{\eta}} {\cal{C}}_{\eta}  \big(n_{\rm{B}} (E_{0}^{\ast })\big)/h$,
which does not depend on the index $\sigma $ with dropping the index $yx$ for convenience. This gives $ G_{\sigma}^{yx } =  L^{\prime}_{11}{\cal{N}}_{0 \sigma  } $
and $ K_{\sigma}^{yx } = (1/T)(L_{22}^{\prime} - {L_{21}^{\prime} L_{12}^{\prime}}/{L_{11}^{\prime}}) {\cal{N}}_{0 \sigma } $.
Consequently,
\begin{subequations}
\begin{eqnarray}
G_{\rm{AF}}^{yx } &=& {L^{\prime}}_{11} ({\cal{N}}_{0 \uparrow }+ {\cal{N}}_{0 \downarrow }),   
\label{eqn:G_magnonHall77}  \\
    K_{\rm{AF}}^{yx } &=& (1/T) \Big( L_{22}^{\prime} - \frac{L_{21}^{\prime} L_{12}^{\prime}}{L_{11}^{\prime}}   \Big) 
({\cal{N}}_{0 \uparrow }+ {\cal{N}}_{0 \downarrow }).
  \label{eqn:K_magnonHall2}
\end{eqnarray}
\end{subequations}
The vanishing of the total Chern number \cite{Z2topo,TIreview,TIreview2} [Eq. (\ref{eqn:ChernPlus})], ${\cal{N}}_{0 \uparrow }+ {\cal{N}}_{0 \downarrow } =0$, results in 
\begin{eqnarray}
G_{\rm{AF}}^{yx } = 0,  \  \  \  \  \  K_{\rm{AF}}^{yx }=0.
  \label{eqn:GKzero}
\end{eqnarray}
Thus in contrast to the magnonic Hall system of FMs \cite{KJD}, the thermomagnetic ratio of the AF $  K_{\rm{AF}}^{yx }/ G_{\rm{AF}}^{yx } $ becomes ill-defined in the sense that the total magnonic spin Hall conductance is zero $G_{\rm{AF}}^{yx } = 0$; the WF law \cite{magnonWF,ReviewMagnon} characterized by the liner-in-$T$ behavior becomes violated since the total magnonic thermal Hall conductance vanishes, i.e. $K_{\rm{AF}}^{yx } = 0$.

Defining the total magnonic spin and heat Hall current densities, 
$   {\cal{J}}_{y} \equiv   \sum_{\sigma }   j_{y \sigma } $ and $   {\cal{J}}_{y}^{Q}  \equiv   \sum_{\sigma }    j_{y \sigma }^{Q}   $, respectively,
Eq. (\ref{eqn:2by2}) is rewritten as
\begin{eqnarray}
%%%%%%%%%%%%%%%%%%%%%%%%%%%%%
\begin{pmatrix}
 \langle  {\cal{J}}_{y}  \rangle   \\     \langle   {\cal{J}}_{y}^{Q}  \rangle
\end{pmatrix}
%%%%%%%%%%%%%%%%%%%%%%%%%%%%%
=
%%%%%%%%%%%%%%%%%%%%%%%%%%%%%
\begin{pmatrix}
 L^{\prime}_{11}({\cal{N}}_{0 \uparrow }+ {\cal{N}}_{0 \downarrow })  & L^{\prime}_{12}({\cal{N}}_{0 \uparrow } - {\cal{N}}_{0 \downarrow })    \\    
 L^{\prime}_{21}({\cal{N}}_{0 \uparrow }- {\cal{N}}_{0 \downarrow })  &  L^{\prime}_{22}({\cal{N}}_{0 \uparrow }+ {\cal{N}}_{0 \downarrow })
\end{pmatrix}
%%%%%%%%%%%%%%%%%%%%%%%%%%%%%
%%%%%%%%%%%%%%%%%%%%%%%%%%%%%
\begin{pmatrix}
  \partial _x  B    \\  -  \frac{\partial _x  T}{T} \nonumber
\end{pmatrix}.
%%%%%%%%%%%%%%%%%%%%%%%%%%%%%
\\   
\label{eqn:2by22222}
\end{eqnarray}
The component of $L^{\prime}_{12}$ represents the magnonic spin Nernst effect in AFs \cite{MagnonNernstAF,MagnonNernstAF2,MagnonNernstExp} where thermal gradients generate {\it{helical}} magnon Hall transport, and consequently, the total magnonic spin Hall current becomes nonzero.
This arises from the opposite magnetic dipole moment $\sigma = \pm 1$ inherent to the N$\acute{{\rm{e}}} $el magnetic order in AF, and the effect is characterized or ensured by the ${\mathbb{Z}}_{2}$ topological invariant \cite{Z2topo,Z2topoHaldane,TIreview,TIreview2,Z2Robust} [Eq. (\ref{eqn:Z2})] ${\cal{Z}}_{0} \equiv ({\cal{N}}_{0 \uparrow } - {\cal{N}}_{0 \downarrow })/2 =  1  $.

The same holds for the reciprocal phenomenon, the magnonic Nernst-Ettinghausen effects \cite{SMreviewMagnon} parametrized by $L^{\prime}_{21}$, while  \cite{QSHE2005,QSHE2006,Z2topo,Z2topoHaldane,Z2SM,TIreview,TIreview2}.
Note that here we refer to the phenomenon described by $L^{\prime}_{11}$ term as `magnonic spin Hall effect' in the bulk AF since it characterizes the magnonic spin Hall conductance $G_{\rm{AF}}^{yx }$ where all magnons subjected to a magnetic field gradient propagate in the same direction and, consequently, the total magnonic spin Hall current becomes zero.
%%%%%%%%%%%%%%%%%%%%%%%
This can be qualitatively understood as follows;
the particle Hall current density for each magnon $ j_{y \sigma }^{\rm{P}} $ is given by $ j_{y \sigma }^{\rm{P}} = j_{y \sigma }/(\sigma  g \mu _{\rm{B}})$,
and Eqs. (\ref{eqn:2by2}) and (\ref{eqn:spectrum222}) provide (see also Table \ref{tab:table1})
\begin{eqnarray}
  \langle  j_{y \sigma }^{\rm{P}} \rangle = \sigma   {\cal{N}}_{0 \sigma } \frac{{L^{\prime}}_{11}}{g \mu _{\rm{B}}} \partial _x  B  
-   {\cal{N}}_{0 \sigma } \frac{{L^{\prime}}_{12}}{g \mu _{\rm{B}}}  \frac{\partial _x  T}{T}.
\label{eqn:ParticleHall}
\end{eqnarray}
Since $ {\cal{N}}_{0 \sigma } = \sigma $, i.e., $  {\cal{N}}_{0 \uparrow } = -{\cal{N}}_{0 \downarrow } =1 $,
it shows that thermal gradient generates {\it{helical}} magnon Hall transport in the topological bulk AF where up and down magnons flow in  opposite $y$ direction, while all magnons subjected to the magnetic field gradient flow in the same $y$ direction  because of the relation $ \sigma  {\cal{N}}_{0 \sigma }  = 1$.
This is in contrast to the topologically trivial bulk AF (Sec. \ref{sec:trivial}) where the magnetic field gradient working as a driving force $F_{\sigma} \propto  \sigma g \mu _{\rm{B}} $
produces helical magnon currents.
The difference arises from each topological integer \cite{QSHE2005,QSHE2006,Z2topo,Z2topoHaldane,Z2SM,TIreview,TIreview2}, i.e., the Chern number 
$  {\cal{N}}_{0 \sigma } = \sigma  $ which leads to $  \sigma  {\cal{N}}_{0 \sigma } =1 $.
%%%%%%%%%%%%%
Note that each magnon by itself carries spin $G_{\sigma}^{yx } \not=0$ and heat $K_{\sigma}^{yx } \not=0$, and 
each mode respectively satisfies the same WF law \cite{KJD} $K_{\sigma}^{yx }/G_{\sigma}^{yx } = [k_{\rm{B}}/(g \mu _{\rm{B}})]^2 T$ as the `quantum' Hall system of ferromagnetic magnons.
However, due to the relation ${\cal{N}}_{0 \uparrow } +{\cal{N}}_{0 \downarrow }=0 $ between each topological integer \cite{QSHE2005,QSHE2006,Z2topo,Z2topoHaldane,Z2SM,TIreview,TIreview2}, magnonic spin and thermal Hall effects in the bulk represented by Eqs. (\ref{eqn:G_magnonHall77}) and (\ref{eqn:K_magnonHall2}), respectively, are prohibited in the topological bulk AF, while the magnonic Nernst-Ettinghausen effects \cite{SMreviewMagnon} shown by the off-diagonals in Eq. (\ref{eqn:2by22222}), $L^{\prime}_{12} $ and $L^{\prime}_{21} $ terms, are characterized or ensured by the ${\mathbb{Z}}_{2}$ topological number ${\cal{Z}}_{0} $
defined in Eq. (\ref{eqn:Z2}).
Thermomagnetic properties of such magnon transport in the topological bulk \cite{MookPrivate} AF are summarized in Table \ref{tab:table1}.

Lastly, regarding the linear response to magnetic field gradient, {\it{e.g.}}, the total magnonic spin Hall conductance $G_{\rm{AF}}^{yx }$ or $ L^{\prime}_{11} $ term, we remark that in Eqs. (\ref{eqn:G_magnonHall77}) and (\ref{eqn:2by22222}), we may still work under the assumption that the relation 
 ${\cal{N}}_{0 \uparrow }+ {\cal{N}}_{0 \downarrow } = 0$ between each topological integer \cite{QSHE2005,QSHE2006,Z2topo,Z2topoHaldane,Z2SM,TIreview,TIreview2} is valid since, just for a perturbative driving force, we assume a (negligibly) small magnetic field gradient that does not disturb the cyclotron motion of magnons; thanks to the spin anisotropy-induced energy gap, the energy spectrum is not affected at all and each edge magnon state remains unchanged thereby ensuring the relation ${\cal{N}}_{0 \uparrow }+ {\cal{N}}_{0 \downarrow } = 0$ between topological integer, i.e., ${\cal{N}}_{0 \sigma } = \sigma $.
Recall that in Sec. \ref{subsec:MQSHE}, we have seen that the cyclotron motion induced by the AC effect leads to the helical edge magnon state characterized by the nonzero Chern number  $ {\cal{N}}_{0 \sigma } = \sigma  $.
Therefore, as long as each magnon type performs a cyclotron motion, the relation remains unchanged. \footnote{The expression, Eqs. (\ref{eqn:G_magnonHall77}) and (\ref{eqn:2by22222}), itself is valid in any case.}

%%%%%%%%%%%%%%%%%%%%%
\section{Estimates for experiments}
\label{sec:experimentAF}
%%%%%%%%%%%%%%%%%%%%%

Observation of spin-wave spin currents \cite{spinwave,WeesNatPhys}, thermal Hall effect of magnons \cite{onose}, magnon planar Hall effect \cite{MagnonHallEffectWees}, 
Snell's law for spin-wave \cite{Snell_Exp,Snell2magnon}, and electrically-induced AC effect \cite{casher,Mignani,magnon2,KKPD,ACatom,KJD} on a magnonic system \cite{ACspinwave} has been reported.
Recently, measurement of magnonic spin conductance \cite{MagnonHallEffectWees} has been reported in Ref. [\onlinecite{MagnonG}] and thermal generation of spin currents in AFs has been established experimentally in Ref. [\onlinecite{SekiAF}] using the spin Seebeck effect \cite{uchidainsulator,ishe,ohnuma,adachi,adachiphonon,xiao2010,OnsagerExperiment,Peltier}, with the subsequent report \cite{MagnonNernstExp,MagnonNernstAF,MagnonNernstAF2} of magnonic spin Nernst effect in AFs.
Moreover, on top of Brillouin light scattering spectroscopy \cite{demokritovReport,demokritov,MagnonPhonon}, using infrared camera, the real-time observation of spin-wave propagation is now possible and Ref. [\onlinecite{Camera2}] reported the observation of magnon Hall-like effect \cite{MagnonHallEffectWees}.

Therefore, we can expect that the observations of the magnonic WF law in the topologically trivial bulk AF and the magnonic QSHE (helical edge magnons) in the topological AF are now within experimental reach \cite{YacobyChemical,KentBEC,GeimSTM,STM_Egradient,ExpPSI,NontopoSurfaceSpinwave} via measurement schemes proposed in Ref. [\onlinecite{KJD}]. 
%%%%%%%
The considered electric field with the constant gradient can be realized by an electric skew-harmonic potential \cite{KJD} and, while being challenging, it may be realized by STM tips \cite{GeimSTM,STM_Egradient}. 
The resulting magnetization gradient from the applied thermal gradient plays a role of an effective magnetic field gradient and works as a nonequilibrium magnonic spin chemical potential \cite{SilsbeeMagnetization,Basso,Basso2,Basso3,MagnonChemicalWees} that has been established experimentally in Ref. [\onlinecite{YacobyChemical}].

For an estimate, we assume the following experiment parameter values \cite{CrOexp,CrOexp2} for ${\rm{Cr}}_2{\rm{O}}_3$, 
$J=15 $meV, ${\cal{K}}=0.03 $meV, $S=3/2$, $g=2$, ${\mathcal{E}} = 1$V/nm$^2$, and $a = 0.5$nm.
This provides the Landau gap $ \hbar \omega _c  = 1 \mu $eV and $ l_{\rm{{\mathcal{E}}}}= 0.7 \mu $m [Eqs. (\ref{omega_c}) and (\ref{l_E})], with which the magnonic QSHE and helical edge magnons could be observed at $T \lesssim 10$mK.
At these low temperatures, effects of magnon-magnon and magnon-phonon interactions can be expected to become negligible \cite{magnonWF,adachiphonon,Tmagnonphonon}.
%%%%%%%%%%%%%%%%%%
An alternative platform to look for topological magnon Hall effects would be skyrmion-like lattices of AFs \cite{AFskyrmion,AFskyrmion2,MookJrAFskyrmion} with Dzyaloshinskii-Moriya (DM) \cite{DM,DM2,DM3} interaction where the N$\acute{{\rm{e}}} $el order parameter varies slowly compared to the typical wavelength of magnons (i.e., spin-waves).
In Ref. [\onlinecite{KevinHallEffect}] (see Appendix \ref{sec:ACvsDM} for details), we have seen that the low-energy magnetic excitations in the skyrmion lattice are magnons 
and the DM interaction \cite{Mook2,Lifa,katsura2} produces intrinsically a vector potential analogous to ${\mathbf{A}}_{\rm{m}}$ which reduces to the same form as Eq. (\ref{HamiltonianLL}).
Assuming experimental parameter values \cite{SkyrmionReviewNagaosa,SkyrmionExpTokura,SkyrmionTheory} (see Ref. [\onlinecite{KJD}] for details),
Landau gaps on the order of a few meVs could be reached.
Since the Hamiltonian for an AF in a skyrmion-like lattice where the N$\acute{{\rm{e}}} $el  order varies slowly (compared to the typical wavelength of spin-waves) also reduces to the qualitatively same form as Eq. (\ref{HamiltonianLL}), we expect that the topological magnon Hall effects could be observed at $T \lesssim {\cal{O}}(10) $K in such AFs \cite{AFskyrmion,AFskyrmion2,MookJrAFskyrmion}.
The temperature, however, should be low enough to make spin-phonon and magnon-magnon contributions negligible \cite{magnonWF,adachiphonon,Tmagnonphonon}.
%%%%%%%%%%%%%%%%%%%%%
As to the magnonic WF law in the topologically trivial bulk AFs, 
the energy gap amounts to $\Delta = 4$meV and thus the magnonic WF law may be observed at $ T =  40$K ($k_{\rm{B}} T = \Delta $).
However, again, the temperature should be low enough \cite{PhononWF} to make spin-phonon and magnon-magnon contributions negligible \cite{magnonWF,adachiphonon,Tmagnonphonon}.
Therefore we expect that the effect becomes observable at low temperature $T \lesssim {\cal{O}}(1) $K. 
\footnote{ Ref. [\onlinecite{Tmagnonphonon}] reported measurements in a magnet  at  low temperature $T \lesssim {\cal{O}}(1) $K which showed that the exponent of the phonon thermal conductance is larger than that of magnons in terms of temperature. This indicates that in terms of thermal conductance, the effects of phonons die out more quickly than magnons at decreasing temperature.}

Given these estimates, we conclude that the observations of the magnonic and topological phenomena in AFs as proposed in this work,
while being challenging, seem within experimental reach \cite{KentPrivate}.

%%%%%%%%%%%%%
\section{Summary}
\label{sec:sum}
%%%%%%%%%%%%%

Under the assumption that the spin along the $z$ direction remains a good quantum number, we have studied thermomagnetic properties of helical transport of magnons with the opposite magnetic dipole moment inherent to the N$\acute{{\rm{e}}} $el order both in topologically trivial and nontrivial bulk AFs.
Since the quantum-mechanical dynamics of magnons in the insulating AF is described as the combination of  independent copies of that in FMs, we found that both topologically trivial magnets satisfy the same magnonic WF law, exhibiting a linear-in-$T$ behavior at sufficiently low temperatures, while the law becomes violated in the topological bulk AF due to the topological invariant that helical edge magnon states bring about.
%%%%%%%%
In the electric field gradient-induced AC effect, up and down magnons form the same Landau energy level and perform cyclotron motion with the same frequency but in opposite directions giving rise to helical edge magnon states, i.e., QSHE of edge magnons, and the AF becomes characterized by the ${\mathbb{Z}}_{2}$ topological number consisting of the Chern integer that each edge state brings about and the AF can be identified as a 
bosonic version of a TI. 
%%%%%%%%
In the almost flat band inherent to the electrically-induced topological AF, the magnonic spin and thermal Hall effects of bulk magnons are prohibited by the topological integer, while the Nernst-Ettinghausen effects are ensured by the ${\mathbb{Z}}_{2}$ topological invariant. The relation between each topological integer is robust against external perturbation as long as magnons can perform cyclotron motion giving the helical edge magnon states.

Finally, it would be interesting to test our predictions experimentally.

%%%%%%%%%%%%%%%%%%%
\section{Discussion}
\label{sec:discussion}

To conclude a few comments on our approach are in order. Instead of deriving the helical edge states directly from the spin Hamiltonian in the presence of electric fields as done previously  for FMs \cite{KJD}, 
here we first derive  the magnon approximation of the spin Hamiltonian in the continuum limit and then introduce the AC phase. The resulting Hamiltonian with quadratic magnon dispersion is then
analyzed numerically by introducing the corresponding TBR. Throughout this paper we have thus restricted our consideration to  AFs where within the long wave-length approximation the dispersion becomes gapped and parabolic, and the dynamics of magnons in the AF can be described as the combination of two independent copies of the dynamics of magnons in a FM for each mode $\sigma =\pm 1$. The helical edge states we found in this approximation are topologically stable and thus 
their emergence does not depend on the microscopic details as long as the gap remains open.
Still, a general treatment of AFs (beyond the parabolic dispersion regime, on different lattices, {\it{e.g.}}, on a triangular spin lattice in the presence of frustration {\it{etc.}})  remains an open issue and deserves further study.

%Dear Daniel,
%in the paragraph below, I tried to reply to the following Referee's comments (i). For this, I mentioned what could happne to the $\sigma$-dependence of the TBR when we apply laser (i.e., Floquet, while we did not use the terminology to keep it secret) since I afraid that otherwise the Referee might say ``trivial''.
%%%%%%%%%%%%%%%%%%%%%%
%(i) The section III.C adds nothing compared to the ferromagnetic case of Ref. 46 but the fact that magnons can have up or down spin, which does not require to repeat all the calculations and numerical results known from Ref. 46. Up to the spin dependence, Eq. (23) of the manuscript is identical to Eq. (2) of Ref. 46.
%In my opinion, the best solution would be to remove III.C from the manuscript and instead refer to Ref. 46 and additionally comment on the differences to the ferromagnetic case. This would avoid repetitions and allow to focus on the differences between the ferromagnetic and the antiferromagnetic case.
%%%%%%%%%%%%%%%%%%%%%%

Lastly we remark that due to the opposite magnetic dipole moments of up and down magnons associated with the magnetic N$\acute{{\rm{e}}} $el order in AFs, the $\sigma$-dependence is simply added to the TBR Eq. (\ref{magnon_hopping}). 
This $\sigma $-dependence, while being a small theoretical difference from the FMs \cite{KJD}, produces qualitatively new phenomena in AFs such as helical edge magnon states and the violation of the magnonic WF law \cite{KJD}. 
%Sec. \ref{subsec:TopoHallBulk}).   
We stress  that this simplicity of the $\sigma$-dependence is specific to the time-independent case considered here. In contrast, when the electric field becomes time-dependent ({\it{e.g.}}, in the presence of laser pulses)
%Applying laser fields, 
the AC gauge potential becomes also time-dependent and the $\sigma$-dependence could be controlled or even vanish for some ac electric fields. This opens up a new control
on the topological phase. It will thus be interesting to study time-dependent effects in these systems in more detail.

\begin{acknowledgments}
We (KN, JK, and DL) acknowledge support by the Swiss National Science Foundation and the NCCR QSIT. One of the authors (SKK) was supported by the Army Research Office under Contract No. W911NF-14-1-0016. We would like to thank A. Mook, V. Zyuzin, A. Zyuzin, H. Katsura, K. Totsuka, K. Usami, J. Shan, C. Schrade, and Y. Tserkovnyak for helpful discussions.
\end{acknowledgments}

\appendix

%%%%%%%%%%%%%%%%%
 \section{Magnons in AFs}
\label{sec:Mchirality}
%%%%%%%%%%%%%%%%%

In this Appendix, we provide some details of the straightforward treatment of AFs \cite{altland,RKuboAF,AndersonAF}  showing that their N$\acute{{\rm{e}}} $el order  provides up and down magnons, i.e., `magnetically charged' bosonic quasiparticles carrying opposite magnetic dipole moments $\sigma  g\mu_{\rm{B}}  {\bf e}_z$.
An external magnetic field $ {\mathbf{B}} = B {\mathbf{e}}_z  $ couples with spins via the Zeeman interaction given by
${\cal{H}}_{B} =  - g\mu_{\rm{B}} B \sum_{l} S_l^z  $.
Assuming spins in the AF form the N$\acute{{\rm{e}}} $el order along  $z$ direction, and using the sublattice-dependent Holstein-Primakoff \cite{HP,altland,MagnonNernstAF,MagnonNernstAF2} transformation,
$S_{i{\rm{A}}}^z = S - a_i^\dagger a_i$, $S_{j{\rm{B}}}^z = - S + b_j^\dagger b_j$, with $[a_{i}, a_{j}^{\dagger }] = \delta_{i,j}$ and $[b_{i}, b_{j}^{\dagger }] = \delta_{i,j}$, we find for the $z$ component of the total spin \cite{MagnonNernstAF}
$  S^z \equiv  \sum_{l} S_l^z = \sum_{i} (S_{i{\rm{A}}}^z  + S_{i{\rm{B}}}^z) = \sum_{i}  (- a_i^\dagger a_i + b_i^\dagger b_i) $.
After Fourier transformation $ S^z =   \sum_{\mathbf{k}} (-a_{\mathbf{k}}^{\dagger } a_{\mathbf{k}} + b_{\mathbf{k}}^{\dagger } b_{\mathbf{k}}) $,
the Hamiltonian becomes 
${\cal{H}}_{B} =  g\mu_{\rm{B}} B \sum_{\mathbf{k}} (a_{\mathbf{k}}^{\dagger } a_{\mathbf{k}} - b_{\mathbf{k}}^{\dagger } b_{\mathbf{k}}) $.
Using a Bogoliubov transformation
\begin{eqnarray}
%%%%%%%%%%%%%%%%%%%%%%%%%%%%%
\begin{pmatrix}
  a_{\mathbf{k}}^{\dagger }   \\     b_{\mathbf{k}}
\end{pmatrix}
=
\cal{M}
%%%%%%%%%%%%%%%%%%%%%%%%%%%%%
\begin{pmatrix}
  {\cal{A}}_{\mathbf{k}}^{\dagger }   \\     {\cal{B}}_{\mathbf{k}}
\end{pmatrix},
%%%%%%%%%%%%%%%%%%%%%
%%%%%%%%%%%%%%%%%%%%%
\begin{pmatrix}
  a_{\mathbf{k}}   \\     b_{\mathbf{k}}^{\dagger }
\end{pmatrix}
%%%%%%%%%%%%%%%%%%%%%%%%%%%%%
=
\cal{M}
%%%%%%%%%%%%%%%%%%%%%%%%%%%%%
\begin{pmatrix}
  {\cal{A}}_{\mathbf{k}}   \\     {\cal{B}}_{\mathbf{k}}^{\dagger }
\end{pmatrix}
%%%%%%%%%%%%%%%%%%%%%%%%%%%%%
\label{eqn:2by2AA}
\end{eqnarray}
with the coefficient matrix $\cal{M}$ defined by 
\begin{eqnarray}
%%%%%%%%%%%%%%%%%%%%%%%%%%%%%
\cal{M}
=
%%%%%%%%%%%%%%%%%%%%%%%%%%%%%
\begin{pmatrix}
 {\rm{cosh}}\vartheta _{\mathbf{k}}  &   -  {\rm{sinh}}\vartheta _{\mathbf{k}}  \\    
  -  {\rm{sinh}}\vartheta _{\mathbf{k}}    &  {\rm{cosh}}\vartheta _{\mathbf{k}}
\end{pmatrix},
%%%%%%%%%%%%%%%%%%%%%%%%%%%%%
\label{eqn:2by2MMM}
\end{eqnarray}
the Hamiltonian $\cal{H}$ in the main text becomes diagonal \cite{altland,RKuboAF,AndersonAF},
$ {\cal{H}} =  \sum_{\mathbf{k}}  \hbar \omega _{\mathbf{k}} ({\cal{A}}_{\mathbf{k}}^{\dagger } {\cal{A}}_{\mathbf{k}} + {\cal{B}}_{\mathbf{k}}^{\dagger } {\cal{B}}_{\mathbf{k}})$, in terms of Bogoliubov quasiparticle operators, ${\cal{A}}_{\mathbf{k}}$ and ${\cal{B}}_{\mathbf{k}}$, satisfying  bosonic commutation relations $ [{\cal{A}}_{\mathbf{k}}, {\cal{A}}_{{\mathbf{k^{\prime}}}}^{\dagger }]= \delta_{{\mathbf{k}}, {\mathbf{k^{\prime}}}}  $ and $ [{\cal{B}}_{\mathbf{k}}, {\cal{B}}_{{\mathbf{k^{\prime}}}}^{\dagger }]= \delta_{{\mathbf{k}}, {\mathbf{k^{\prime}}}}  $, where $ {\rm{tanh}} (2 \vartheta _{\mathbf{k}}) = \gamma _{\mathbf{k}}/(1+\kappa) $,
 $\gamma _{\mathbf{k}} = (1/\rho ) \sum_{m=1}^{\rho } {\rm{e}}^{- i {\mathbf{k}}\cdot {\boldsymbol{\delta }}_{m} }  $, the coordination number $ \rho =2d$, and ${\boldsymbol{\delta }}_{m}$ the relative coordinate vector that connects the nearest neighboring sites. The $z$ component of the total spin is rewritten as 
$  S^z  = \sum_{\mathbf{k}} (-a_{\mathbf{k}}^{\dagger } a_{\mathbf{k}} + b_{\mathbf{k}}^{\dagger } b_{\mathbf{k}}) = \sum_{\mathbf{k}}(-{\cal{A}}_{\mathbf{k}}^{\dagger } {\cal{A}}_{\mathbf{k}} + {\cal{B}}_{\mathbf{k}}^{\dagger } {\cal{B}}_{\mathbf{k}}) $
and thereby 
${\cal{H}}_{B} =  g\mu_{\rm{B}} B \sum_{\mathbf{k}} (a_{\mathbf{k}}^{\dagger } a_{\mathbf{k}} - b_{\mathbf{k}}^{\dagger } b_{\mathbf{k}}) = g\mu_{\rm{B}} B \sum_{\mathbf{k}}({\cal{A}}_{\mathbf{k}}^{\dagger } {\cal{A}}_{\mathbf{k}} - {\cal{B}}_{\mathbf{k}}^{\dagger } {\cal{B}}_{\mathbf{k}}) $.
%%%%%%%%%%%%%%%%%%%%%%%%
Therefore, it can be seen that the ${\cal{A}}$- (${\cal{B}}$-) magnon carries $\sigma = -1$ ($+1$) spin angular momentum along the $z$ direction and can be identified with down and up magnons, respectively.

In the absence of the magnetic field, $B=0$, these up and down magnons are degenerate and the energy dispersion \cite{RKuboAF,AndersonAF} is given by
$ \hbar \omega _{{\mathbf{k}}} = 2J dS \sqrt{(1+\kappa)^2 - \gamma _{\mathbf{k}}^2}   $.
Within the long wave-length approximation, it becomes $  \gamma _{\mathbf{k}}^2= 1- (ak)^2/d   $ for $ \mid  {\mathbf{k}} \mid  = k $ and thereby assuming a spin anisotropy \cite{CrOexp,CrOexp2} at low temperature, the dispersion becomes parabolic in terms of $k$ and reduces to the form $\hbar \omega _{{\mathbf{k}}} = Dk^2 + \Delta$ with $   D = JSa^2/\sqrt{\kappa^2 + 2\kappa}  $ which is used in the main text.
Note that the dispersion becomes linear in terms of $k$, $\hbar \omega _{{\mathbf{k}}} \propto k $, when there is no spin anisotropy $ {\cal{K}} =0$, i.e., $ \kappa =0$.

Lastly we remark that the $z$ component of spins is a good quantum number \cite{Z2Robust} of our system, which commutes with the original spin Hamiltonian [Eq. (\ref{eqn:H})].  
Therefore regardless of the analytical approach taken, {\it{e.g.}}, noninteracting magnon picture \cite{RKuboAF,AndersonAF} using the Holstein-Primakoff transformation we adopted throughout this work, the excitations should have a  well-defined spin $z$ component. The Hamiltonian and the spin $z$ component are simultaneously diagonalizable.  Therefore it can be expected that, apart from any magnon picture, two well-defined opposite spin modes and the helical nature of the resultant edge spin modes should survive in any case at sufficiently low temperatures where phonons die out \cite{adachiphonon,Tmagnonphonon}.

%%%%%%%%%%%%%%%%%%%%%%%%%%
\section{Boltzmann equation for magnons}
\label{sec:Boltzmann}
%%%%%%%%%%%%%%%%%%%%%%%%%%

In this Appendix, we provide some details of the straightforward calculation for the Onsager coefficients $L_{i j \sigma }$ in the topologically trivial bulk AF.
Assuming the system is slightly out of equilibrium and using the Boltzmann transport equation \cite{Basso,Basso2,Basso3} given in Ref. [\onlinecite{mahan}], the Bose-distribution function of magnons $f_{{\mathbf{k}}\sigma } $ becomes  
$f_{{\mathbf{k}}\sigma } = f^0_{{\mathbf{k}}\sigma }  + g_{{\mathbf{k}}\sigma } $ where
$ f^0_{{\mathbf{k}}\sigma } = ({\rm{e}}^{\beta  \epsilon _{{{\mathbf{k}}\sigma }}}-1)^{-1}   $ with $ \epsilon _{{{\mathbf{k}}\sigma }} = \hbar \omega _{{\mathbf{k}}\sigma }  $ is the equilibrium distribution while the deviation from equilibrium 
$ g_{{\mathbf{k}}\sigma } =  f_{{\mathbf{k}}\sigma } - f^0_{{\mathbf{k}}\sigma }  $ is given by 
$  g_{{\mathbf{k}}\sigma } = \tau {\mathbf{v}}_{{\mathbf{k}}} \cdot [-\sigma  g \mu_{\rm{B}} {\mathbf{\nabla }}B + (\epsilon _{{{\mathbf{k}}\sigma }}/T){\mathbf{\nabla }}T] (\partial  f^0_{{\mathbf{k}}\sigma }/\partial \epsilon _{{{\mathbf{k}}\sigma }}) $ within the liner response regime, where  
$ {\mathbf{v}}_{{\mathbf{k}}} = \partial \epsilon _{{\mathbf{k}}\sigma }/\partial \hbar {\mathbf{k}}$ is the velocity and $\tau$ a phenomenologically introduced relaxation time of magnons, mainly due to nonmagnetic impurity scatterings and thereby we may assume it to be a constant at low temperature.
The resulting particle, spin, and heat currents for each magnon mode,
$ {\mathbf{j}}_{\sigma }^{\rm{P}} $, $ {\mathbf{j}}_{\sigma } $, $ {\mathbf{j}}_{\sigma }^{Q} $, respectively, are given by 
$ {\mathbf{j}}_{\sigma }^{\rm{P}} = \int [d^3{\mathbf{k}}/(2 \pi)^3]    {\mathbf{v}}_{{\mathbf{k}}} g_{{\mathbf{k}}\sigma }   $,
$ {\mathbf{j}}_{\sigma } = \int [d^3{\mathbf{k}}/(2 \pi)^3]   \sigma  g \mu_{\rm{B}} {\mathbf{v}}_{{\mathbf{k}}} g_{{\mathbf{k}}\sigma }   $,
$ {\mathbf{j}}_{\sigma }^{Q} = \int [d^3{\mathbf{k}}/(2 \pi)^3] \epsilon _{{\mathbf{k}}\sigma }   {\mathbf{v}}_{{\mathbf{k}}} g_{{\mathbf{k}}\sigma } $.
Assuming spatial isotropy $ |k_x  | =  | k_y  | = | k_z  |   $ for ${\mathbf{k}} = (k_x, k_y, k_z)$ and performing the Gaussian integrals, the Onsager coefficients $L_{i j \sigma }$ shown in the main text are obtained.

%%%%%%%%%%%%%%%%%%%%%%%%%
\section{Cyclotron motion of magnons}
\label{sec:CyclotronMotion}
%%%%%%%%%%%%%%%%%%%%%%%%%

In this Appendix, we provide for completeness some details of the straightforward calculation showing that the dynamics of magnons with the opposite magnetic dipole moments $ \sigma  g \mu _{\rm{B}}{\mathbf{e}}_z$ are identical except that the resulting  chirality of the magnon propagation becomes opposite [Fig. \ref{fig:HelicalAFChiralFM} (b)].
Using the correspondence explained in the main text, the calculation becomes analogous to the one for electrons \cite{mahan,Ezawa}, and especially it parallels  the one for ferromagnetic magnons given in Ref. [\onlinecite{KJD}] except that we have now  two magnon modes with opposite magnetic dipole moment ($\sigma =\pm 1$).
%%%%%%%%%%%%%%%%%%%%%%%
Introducing operators analogous to a covariant momentum $ {\hat{{\mathbf{\Pi}}} }_{\sigma } \equiv   \hat{{\mathbf{p}} }  +  \sigma g \mu _{\rm{B}}{\mathbf{A}}_{\rm{m}} /c =(\Pi_{x \sigma }, \Pi_{y \sigma }) $, which satisfy $[\Pi_{x \sigma }, \Pi_{y \sigma }]= - i \sigma  \hbar ^2/ l_{\rm{{\mathcal{E}}}}^2  $, and dropping the irrelevant constant, the Hamiltonian for each mode can be rewritten as 
$ {\cal{H}}_{{\rm{m}}\sigma }  = (\Pi_{x\sigma }^2 + \Pi_{y\sigma }^2)/2m $.
Next, introducing  the operators $ A_{\sigma } \equiv     l_{\rm{{\mathcal{E}}}} (\Pi_{x \sigma } - i \sigma  \Pi_{y \sigma })/\sqrt{2} \hbar  $ and $A_{\sigma }^{\dagger }\equiv     l_{\rm{{\mathcal{E}}}} (\Pi_{x \sigma } + i \sigma  \Pi_{y \sigma })/\sqrt{2} \hbar $, which satisfy bosonic commutation relations, 
$ [A_{\sigma }, A_{\sigma }^{\dagger }] = 1$ with the remaining commutators vanishing, the Hamiltonian becomes 
 $ {\cal{H}}_{{\rm{m}}\sigma } =  \hbar  \omega _c  (A_{\sigma }^{\dagger } A_{\sigma } + 1/2)$.
Indeed, introducing \cite{Ezawa} the guiding-center coordinate by $ X_{\sigma } =  x  -  \sigma l_{\rm{{\mathcal{E}}}}^2 \Pi_{y\sigma }/\hbar   $ and $ Y_{\sigma } = y  +  \sigma  l_{\rm{{\mathcal{E}}}}^2 \Pi_{x \sigma }/\hbar  $, which satisfy $[X_{\sigma }, Y_{\sigma }]=  i \sigma  l_{\rm{{\mathcal{E}}}}^2$ with $ dX_{\sigma }/dt = dY_{\sigma }/dt=0 $ 
indicating that the drift velocity vanishes in the absence of the magnetic field gradient,  
the time-evolution of the relative coordinate
 ${\mathbf{R}}_{{\rm{{\mathcal{E}}}}\sigma } =({\cal{R}}_{x \sigma }, {\cal{R}}_{y \sigma }) \equiv  (-  l_{\rm{{\mathcal{E}}}}^2 \Pi_{y \sigma }/\hbar,  l_{\rm{{\mathcal{E}}}}^2 \Pi_{x \sigma }/\hbar)$ becomes 
$  d({\cal{R}}_{x \sigma } + i {\cal{R}}_{y \sigma }) /dt =  i \sigma  \omega _c  ({\cal{R}}_{x \sigma } + i {\cal{R}}_{y \sigma })$.
Thus in the presence of an electric field gradient, two magnons form the same Landau level and perform cyclotron motion with the same frequency, but propagate in  opposite directions due to the opposite magnetic dipole moment $ \sigma  g \mu _{\rm{B}}{\mathbf{e}}_z$ [Fig. \ref{fig:HelicalAFChiralFM} (b)].

%%%%%%%%%%%%%%%%%%%%%%%%%%% 
\section{Landau levels in topological magnets}
\label{sec:ACvsDM}
%%%%%%%%%%%%%%%%%%%%%%%%%%%

In this Appendix, we provide some insight into the magnons in DM interaction-induced skyrmion-like structures where the DM \cite{DM,DM2,DM3} interaction provides \cite{katsura2} an effective AC phase.
In Ref. [\onlinecite{KevinHallEffect}], we have seen that the low-energy magnetic excitations in the skyrmion lattice are magnons and the DM interaction produces a textured equilibrium magnetization that works intrinsically as a vector potential analogous to ${\mathbf{A}}_{\rm{m}}$. 
The Hamiltonian of magnons indeed reduces to the same form of Eq. (\ref{HamiltonianLL}) with the analog of the Landau gauge that produces the Landau energy level [Eq. (\ref{LL})]. Assuming the magnitude of the DM interaction \cite{SkyrmionReviewNagaosa,SkyrmionExpTokura,SkyrmionTheory,DMposition,DMposition2}
$\Gamma _{\rm{DM}}$, the Landau energy level spacing is given by \cite{KevinHallEffect} $ (4JS/\sqrt{3} \pi) (\Gamma _{\rm{DM}}/J)^2   $; see Refs. [\onlinecite{KevinHallEffect,KJD}] for details. Using the correspondence with the Landau energy level spacing by electric field gradient-induced AC effect $\hbar  \omega _{\rm{c}}$ [Eqs. (\ref{LL}) and (\ref{omega_c})], it can be seen to be qualitatively identified with an effective inner electric field gradient \cite{QSHE2016} and the magnitude is estimated by $  {\cal{E}}_{\rm{DM}} = [2 /(\sqrt{3} \pi a^2)] (\hbar c^2/g \mu_{\rm{B}})(\Gamma _{\rm{DM}}/J)^2 \propto  \Gamma _{\rm{DM}}^2 $.
This indicates that the DM interaction produces a slowly-varying textured equilibrium magnetization that provides an effective AC phase and in such a skyrmion-like structure, it works as an effective, fictitious, and intrinsic electric field gradient  ${\cal{E}}_{\rm{DM}} = {\cal{O}}(10^2)$V/nm$^2$ of very large magnitude \cite{KevinHallEffect,katsura2}.
Note that the key to edge magnon states is the vector potential $ {\mathbf{A}}_{\rm{m}}$ that globally satisfies the relation [Eq. (\ref{eqn:correspondenceEB})] ${\mathbf{\nabla }} \times  {\mathbf{A}}_{\rm{m}} = ({{\mathcal{E}}}/{c}) {\mathbf{e}}_z$ where magnons experience the vector potential macroscopically, leading to cyclotron motion.

\bibliography{PumpingRef}

\begin{thebibliography}{137}
\bibitem[{Pho()}]{PhononWF}
\bibinfo{note}{At low temperatures $T\sim 10^{-1}$K, phonon contributions are
  negligible \cite{adachiphonon}, which agrees with the experiment
  \cite{Tmagnonphonon}.}
\end{thebibliography}

\end{document}